\documentclass[usenames,dvipsnames,svgnames]{aastex63}

\usepackage{xcolor}

\usepackage[utf8]{inputenc}
\usepackage{amsmath}
\usepackage{amsfonts}
\usepackage{amssymb}
\usepackage{gensymb}
\usepackage{tabularx}
\usepackage{mathtools}
\usepackage{graphicx}
\usepackage{wrapfig}
\usepackage{color}
\usepackage{enumitem}
\usepackage{placeins}
\usepackage{savesym}
\usepackage{breakurl}
 \def\UrlBreaks{\do\/\do-}

\usepackage[colorinlistoftodos,prependcaption,textsize=footnotesize]{todonotes}

\usepackage{listings}
\usepackage{enumitem}%
\usepackage[T1]{fontenc}
\usepackage{booktabs,tikz}
\usepackage{csquotes}
\usepackage{afterpage}
\usepackage{parskip}
\usepackage{threeparttable}
\usepackage{rotating} %

\graphicspath{
{./figs/}
{../tex-support/}
{./}
{./figs/science/}
{./figs/power/}
{./figs/edl/}
{./figs/history/}
{./figs/communications/}
{./figs/autonomy/}
}

\newif\ifaaspsj
\aaspsjtrue

\newif\ifdeluxetable
\deluxetabletrue

\newif\ifanonymous
\anonymousfalse

\newif\ifmargincomments
\margincommentsfalse
  
\ifmargincomments

\newcommand{\mmmargin}[2]{{\color{red}#1}\marginpar{\color{red}\raggedright\footnotesize [MM]:#2}}
\newcommand{\revmm}[1]{{\color{blue}#1}}
\newcommand{\frtodo}[1]{\todo[inline,color=Apricot]{[fr] #1}}
\newcommand{\revI}[1]{{\textbf{\color{blue}#1}}}
\else

\newcommand{\mmmargin}[2]{#1}
\newcommand{\revmm}[1]{#1}
\newcommand{\frtodo}[1]{}
\fi

\newif\ifshowrevI
\showrevIfalse
\ifshowrevI
\newcommand{\revI}[1]{{\textbf{\color{blue}#1}}}
\else
\newcommand{\revI}[1]{#1}
\fi

\newcommand{\coloredtwopie}[3]{%
\begin{tikzpicture}
 \draw[#3] (0,0) circle (.8ex);\fill[rotate=90-#2,fill=#3] (.8ex,0) arc (0:-#1:.8ex) -- (0,0) -- cycle;
\end{tikzpicture}%
}

\newcommand{\coloredpie}[2]{%
\coloredtwopie{#1}{0}{#2}%
}

\begin{document}

\title{Distributed Instruments for Planetary Surface Science: \\Scientific Opportunities and Technology Feasibility}

\ifanonymous

\author{Authors anonymized for review}

\else
\correspondingauthor{Federico Rossi}
\email{federico.rossi@jpl.nasa.gov}

\author[0000-0002-8091-881X]{Federico Rossi}
\affiliation{Jet Propulsion Laboratory - California Institute of Technology \\
4800 Oak Grove Dr. \\
Pasadena (CA) 91109, USA}

\author{Robert C. Anderson}
\affiliation{Jet Propulsion Laboratory - California Institute of Technology \\
4800 Oak Grove Dr. \\
Pasadena (CA) 91109, USA}

\author[0000-0002-8107-2617]{Saptarshi Bandyopadhyay}
\affiliation{Jet Propulsion Laboratory - California Institute of Technology \\
4800 Oak Grove Dr. \\
Pasadena (CA) 91109, USA}

\author[0000-0001-6106-7645]{Erik Brandon}
\affiliation{Jet Propulsion Laboratory - California Institute of Technology \\
4800 Oak Grove Dr. \\
Pasadena (CA) 91109, USA}

\author[0000-0003-0938-0401]{Ashish Goel}
\affiliation{Jet Propulsion Laboratory - California Institute of Technology \\
4800 Oak Grove Dr. \\
Pasadena (CA) 91109, USA}

\author[0000-0002-0493-237X]{Joshua Vander Hook}
\affiliation{Jet Propulsion Laboratory - California Institute of Technology \\
4800 Oak Grove Dr. \\
Pasadena (CA) 91109, USA}

\author[0000-0002-8022-5319]{Michael Mischna}
\affiliation{Jet Propulsion Laboratory - California Institute of Technology \\
4800 Oak Grove Dr. \\
Pasadena (CA) 91109, USA}

\author[0000-0003-1947-7741]{Michaela Villarreal}
\affiliation{Jet Propulsion Laboratory - California Institute of Technology \\
4800 Oak Grove Dr. \\
Pasadena (CA) 91109, USA}

\author[0000-0002-6521-3256]{Mark Wronkiewicz}
\affiliation{Jet Propulsion Laboratory - California Institute of Technology \\
4800 Oak Grove Dr. \\
Pasadena (CA) 91109, USA}
\fi

\begin{abstract}

In this paper, we assess the scientific promise and technology feasibility of distributed instruments for planetary science. A distributed instrument is an instrument designed to collect spatially and temporally correlated data from multiple networked, geographically distributed point sensors. Distributed instruments are ubiquitous in Earth science, where they are routinely employed for weather and climate science, seismic studies and resource prospecting, and detection of industrial emissions. However, to date, their adoption in planetary surface science has been minimal. It is natural to ask whether this lack of adoption is driven by low potential to address high-priority questions in planetary science; immature technology; or both.
To address this question, we survey high-priority planetary science questions that are uniquely well-suited to distributed instruments. We identify four areas of research where distributed instruments hold promise to unlock answers that are largely inaccessible to monolithic sensors, namely, weather and climate studies of Mars; localization of seismic events on rocky and icy bodies; localization of trace gas emissions, primarily on Mars; and magnetometry studies of internal composition.
Next, we survey enabling technologies for distributed sensors and assess their maturity. We identify sensor placement (including descent and landing on planetary surfaces), power, and instrument autonomy as three key areas requiring further investment to enable future distributed instruments.
Overall, this work shows that distributed instruments hold great promise for planetary science, and paves the way for follow-on studies of future distributed instruments for Solar System in-situ science.

\end{abstract}


\section{Introduction}
\label{ch:intro}

The goal of this paper is to survey the breakthrough scientific opportunities offered by distributed instruments for planetary surface science, and assess their technology feasibility. %

\begin{figure}[h]
 \centering
\includegraphics[width=.85\textwidth]{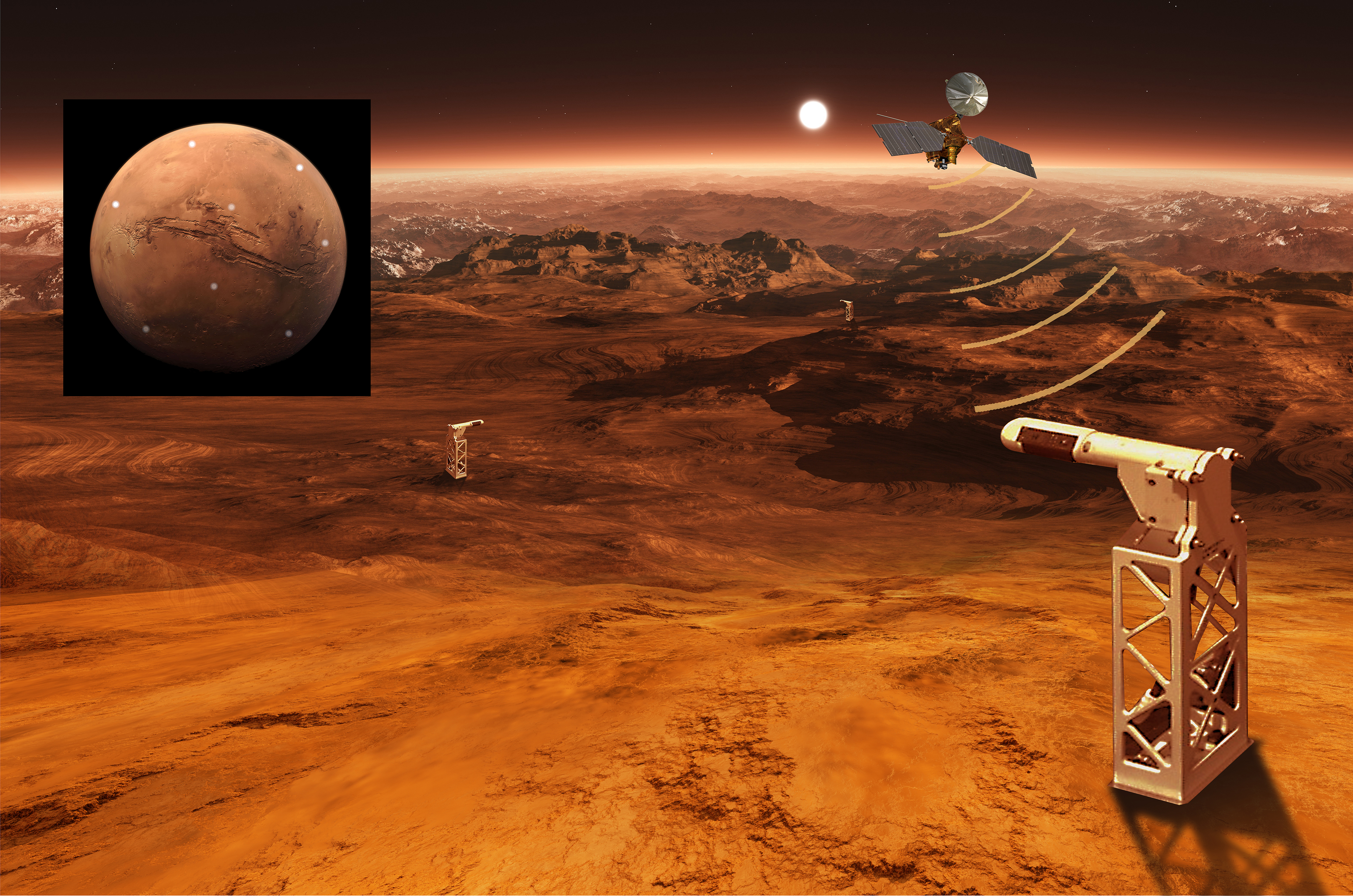}
 \caption{Artist's representation of a notional distributed instrument to study the Martian global climate. A network of sixteen sensing units is distributed across the Martian surface to provide coverage of the entire planet. Each sensing unit contains a weather station, a dust instrument, a camera to observe atmospheric opacity and cloud formation, and a spectrometer. Sensor readings are relayed to Earth by an existing orbiter. The \emph{distributed instrument} allows scientists to study the spatial and temporal correlation of weather and climate patterns (in particular the water, dust, and CO$_2$ cycles) on the planet.}
 \label{fig:intro:di_concept}
 \end{figure}

A distributed instrument is an instrument designed to collect spatially and temporally correlated data from many networked, geographically distributed point sensors (Figure \ref{fig:intro:di_concept}).

A distributed instrument is more than the sum of its sensors; by strategically scheduling measurements across a region to capture data with spatial context, it can provide insight into spatially and temporally correlated phenomena that is impossible to obtain using monolithic rovers or orbiters.

Distributed instruments hold promise to unlock answers to key questions in planetary science that cannot be addressed by point measurements or orbiters alone, including how solar tides influence the atmospheres of bodies like Mars, Venus, and Titan \citep{haberle2014preliminary}, how pressure waves propagate across a planet, the surface locations from where trace gas emissions originate \citep{mischna2020gasdetection}, and what the interior composition of icy bodies is \citep{TitanSeismology,Vance}. The consensus within many scientific disciplines is that, for some fundamental science questions, the output received from a distributed instrument of modest capability would be significantly more valuable than the output from a single sensor with exceptional capabilities \citep{Beaty2012MEPAG,BanfieldEtAl2020MEPAG,Barba2019ICESAG}.
However, the design and development of distributed instruments requires technology advances not only in the design of individual detectors, but also in distributed filtering (to manage the large amount of data produced by distributed instruments), localization and synchronization (to provide spatial and temporal context for the measurements), delay-tolerant and mesh networking (to reliably transmit data to Earth), and entry, descent, and landing (EDL) and swarm guidance (to achieve the desired distribution of the individual sensors).

Today, many individual technologies for distributed instruments are arguably mature or at a tipping point, thanks to recent advances in miniaturization of sensor technology (e.g., \cite{buehler2007electrical}), distributed sensing \citep{tokekar2016sensor}, delay-tolerant networking \citep{burleigh2003dtn}, orbit-based localization \citep{CheungLeeEtAl2019}, and EDL and swarm guidance \citep{Printable,kang2019marsbee,MarsDrop}.

In addition, distributed instruments are routinely used for Earth science, e.g., for earthquake detection \citep{geological2017advanced} and warning \citep{usgs2018revised}, to inform weather forecasts and climate studies \citep{NWS23observationsystems}, and for detection of industrial methane leaks \citep{canary2023cem,chen2023emission}.

Despite this, comparatively few distributed instrument designs have emerged within the planetary science community. In this paper, we take a first step at exploring the science promise and technology feasibility of such distributed instruments by (i) surveying high-priority scientific questions that can be uniquely addressed by distributed instruments with existing or tipping-point technologies; and
(ii) identifying and ranking technologies that should be further developed to ensure the technology feasibility and to maximize the scientific returns of promising distributed instruments.

\subsection{What is a distributed instrument?}

We define a distributed instrument as

\begin{displayquote}
``a collection of \emph{geographically distributed} sensors \revmm{designed to} \emph{strategically} collect \emph{spatially and temporally correlated} readings at known locations.''
\end{displayquote}

The three key characteristics of a distributed instruments are geographical distribution, strategic sampling, and spatial and temporal correlation.

\textbf{Geographically Distributed}: A distributed instrument collects observations from multiple distinct locations through multiple sensing units. Depending on the nature of the instrument and on scientific requirements, the sensing units can be deployed at a local scale (where sensors are separated by tens to hundreds of meters, and cover an area of a few square kilometers), regional scale (where sensors are separated by kilometers to tens of kilometers, and cover areas of thousands to hundreds of thousands of square kilometers), or global scale (where sensor coverage extends to the entire surface of a planetary body).

\textbf{Strategic sampling}: The sensing units act as a single instrument, and their observations are processed to collectively address a science question. This key characteristic of distributed instruments is discussed further in Section \ref{sec:intro:di:sensor-network}.

\textbf{Spatially and Temporally Correlated}: The locations and times of individual sensor readings are well-known, and the sensors themselves are calibrated against each other, allowing scientists to fuse the individual units' readings to characterize the spatial and temporal evolution of the phenomenon under observation.

\subsubsection{Distributed Instruments and Sensor Networks}
\label{sec:intro:di:sensor-network}

Distributed instruments are a subset of sensor networks: that is, all distributed instruments are sensor networks, but not all sensor networks are distributed instruments.
A sensor network is a collection of geographically distributed sensors  (e.g., a set of weather stations). 
A distributed instrument builds on this concept by capturing, \emph{by design}, the spatial and temporal correlation of the readings, acting as a unified set of components collectively addressing a specific science question. 
For a concrete example, a network of weather stations, each independently monitoring atmospheric pressure, is a sensor network but not necessarily a distributed instrument; in contrast, a network of weather stations \revmm{set in carefully chosen locations and sampling atmospheric pressure at agreed-upon times}
 to collectively detect atmospheric tides is a distributed instrument.

\subsubsection{Distributed Instruments and Swarms}

Not all distributed instruments are swarms. ``Swarms'' are loosely defined as a large number of autonomous agents that act based on local information in a way leading to a desired emerging behavior \citep{bss2018swarms}. Typically, individual agents are low-cost and not individually endowed with sophisticated capabilities.

In contrast, a distributed instrument can have as few as two sensing units; each sensing unit can act based on information from all other units, if necessary; and individual units are not necessarily low cost nor unsophisticated. Concretely, two networked seismometers collecting data to localize Mars temblors are a distributed instrument; in contrast, %
a hundred low-cost weather stations individually relaying independent measurements to Earth
\revmm{with no cross-calibration or synchronization} are not  a distributed instrument, \revmm{as they are not designed to accurately capture spatial and temporal correlations}.

\subsection{Contribution and Organization}

The rest of this paper lays out the science case for distributed instruments, and surveys key technologies required to bring their potential to fruition.

Specifically, in Section \ref{ch:history}, we provide a historical perspective of proposed distributed instruments for planetary exploration, with a focus on NASA mission studies.
In Section \ref{ch:sciencecase}, we survey high-priority scientific questions that are uniquely well-suited for distributed instruments, i.e., cannot be addressed with single-point measurements provided by orbiters, landers, or rovers. We further examine, in detail, four promising distributed instrument concepts for atmospheric science, seismology, magnetometry, and trace gas detection and localization.
Section \ref{ch:technologies} provides a broad survey of the state-of-the art in enabling technologies for distributed instruments, including (i) EDL and sensor placement, (ii) localization and synchronization, (iii) communications, (iv) on-board computing and autonomy, (v) power, and (vi) thermal control. Based on the survey, we then highlight which technologies require additional development to bring to reality the instrument concepts in Section \ref{ch:sciencecase}.
Finally, in Section \ref{ch:conclusions} we draw conclusions and lay out recommendations for future research and development.

\section{State of the Art and Historical Perspective}
\label{ch:history}

To date, no distributed instrument has flown to a planetary surface. However, the concept of sensor networks for space exploration (of which distributed instruments are a subset) has been explored through several mission concepts and projects over the last several decades.  In this section, we review some of these efforts; we focus on identifying how recent technological advances have removed significant roadblocks that hindered adoption of such architectures in the past, in particular in the context of planetary surface missions.

\subsection{MESUR}

\begin{figure}[h]
	\centering
	\includegraphics[width=0.5\textwidth]{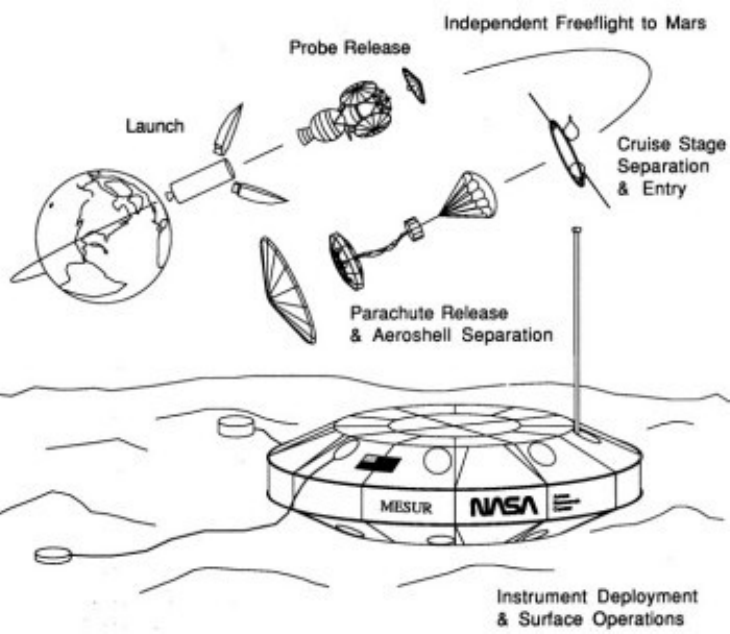}
	\caption{Illustration of the MESUR project's geophysical observation stations (\copyright{} NASA).}
	\label{fig.MESUR}
\end{figure}

Concepts for deploying a network of landers on the Martian surface were being investigated as far back as the early 1990s. The Mars Environmental Survey (MESUR) concept (Figure \ref{fig.MESUR}) was initiated at NASA Ames Research Center, while the detailed design was carried out at NASA JPL  \citep{MESUR,hubbard1992mars}. The goal of the MESUR concept was to deploy sixteen landers spread across the surface of Mars, spanning all latitudes and longitudes.

Concerns regarding a lack of understanding of global circulation patterns in the atmosphere, crustal thickness, state and structure of the mantle, CO\textsubscript{2} and H\textsubscript{2}O deposits in the polar regions, presence of volatiles, organic compounds, soil water inventory, and thermal structure of the upper atmosphere were some of the broad science questions that were identified as being addressable only with the help of multiple networked instruments \citep{MESUR}. In order to meet these objectives, an instrument package was identified comprising: a seismometer; a meteorology package comprising pressure sensor, wind sensor, temperature sensor, sky radiometer and humidity sensor; an Alpha/proton/X-ray spectrometer with a mini rover for transporting the instrument to the desired target; a thermal analyzer (calorimeter); a gas chromatograph; and a descent/surface imager.

The mission concept also included an orbiter for atmospheric sounding and for relaying data from individual landers. The data volume was calculated to be on the order of 100 Mbit/node/day, driven largely by the data requirements of the seismometer. However, it was acknowledged that these requirements could be reduced down to 10 Mbit/node/day, once the seismic environment has been understood well enough to employ event detection and data compression algorithms. 

It was estimated that this network could be established over 5 launches and 3 launch windows. However, given the complexity of each lander module targeting a plethora of science objectives, funding constraints led to the approval of a single lander, initially known as Surface Lander Investigation of Mars (SLIM). The focus of the project shifted to the micro-rover Mars Pathfinder, which was designed to carry and deploy instruments on the Martian surface, and the seismometer was removed from the mission.

\subsection{SensorWeb}

\begin{figure}[h]

	\centering
	\ifaaspsj
	\gridline{
	\fig{SensorWeb_Module.pdf}  {.4\textwidth}{(a) Functional SensorWeb pod \citep{delin2000sensor}.}
	\fig{SensorWeb_Deployment.pdf}{0.4\textwidth}{(b) Proposed planetary surface deployment schemes for the sensor web \citep{delin2000sensor}.}
	}
	\else
	
	\begin{subfigure}[b]{0.4\textwidth}
	\includegraphics[width=\textwidth]{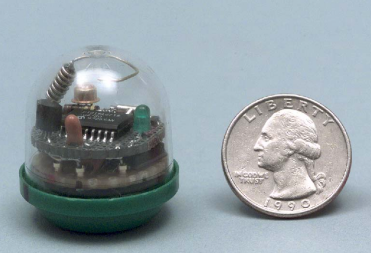}
	\caption{Functional SensorWeb pod \citep{delin2000sensor}.}
	\label{fig.SensorPod}
	\end{subfigure}
	\begin{subfigure}[b]{0.4\textwidth}
		\includegraphics[width=\textwidth]{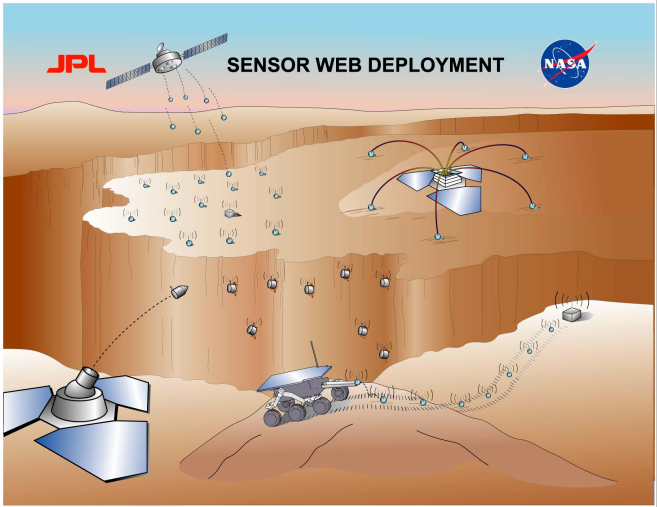}
	\caption{Proposed planetary surface deployment schemes for the sensor web \citep{delin2000sensor}.}
	\label{fig.SensorPodDispersal}	
	\end{subfigure}
	\fi
	
	\caption{NASA JPL's SensorWeb distributed instrument concept}
	\label{fig:sensorweb}
\end{figure}

In the late 1990s, the NASA JPL-developed SensorWeb concept entailed miniaturized, networked sensor modules for a variety of planetary and Earth science applications \citep{delin2000sensor}.
The effort resulted in a commercial spin-off that led to successful deployment of sensor networks for studying, among others, algal blooms in Baja California, biological flourishes in the cryogenic environment of Antarctica, and soil temperature in Huntington Gardens in Pasadena, CA.

The SensorWeb project demonstrated a number of advances in the miniaturization of sensor pods (Figure~\ifaaspsj\ref{fig:sensorweb}(a)~\else\ref{fig.SensorPod}\fi), and in the implementation of multi-hop communication networks. The sensor nodes operated in a dense network, where each sensor relayed its data to all neighbors within communication range, until data reached ``prime'' nodes that would relay all the collected data through a satellite link. 
The architecture resulted in inherent robustness (since no reconfiguration was necessary in the event of node failure or replacement, as long as there was a sufficient density of nodes), and low power usage for communications, since free-space path losses grow with the square of the distance between nodes.  The proposed SensorWeb architecture envisioned heterogeneous sensor nodes with continuously scanning, low-power, high dynamic range sensors for event detection that triggered higher fidelity, power-intensive sensors for precise measurements. Event localization could also be achieved by synthesizing spatio-temporal information from multiple sensors. 

A variety of solutions were considered for dispersal of these sensor pods on planetary surfaces, such as dispersal from an aerial vehicle, projectile deployment from a lander, and placement by a rover while moving across the planetary surface. Figure \ifaaspsj\ref{fig:sensorweb}(b)~\else\ref{fig.SensorPodDispersal}\fi illustrates some of these concepts. 

\begin{figure}
	\centering

\end{figure}

\subsection{Volcano SensorWeb}

\begin{figure}[h]
	\centering
	\includegraphics[width=0.7\textwidth]{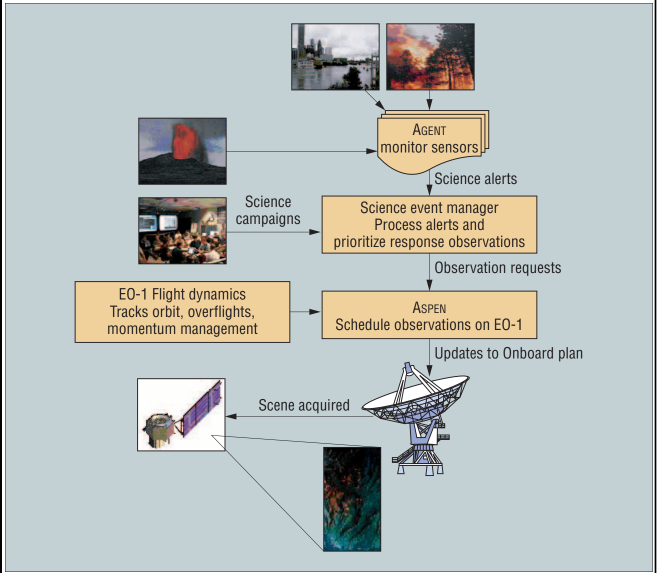}
	\caption{Volcano SensorWeb architecture illustrating how retasking happens upon triggering based on event detection \citep{Chien2005}.}
	\label{fig.ChienSensorWeb}
\end{figure}

In the mid-2000s, NASA JPL developed and implemented an autonomous, Earth-observing sensor web where multiple instruments across remote sensing and in-situ platforms acted in a coordinated manner to provide high-resolution data within hours of critical events such as volcanic eruptions \citep{Chien2005}. Data from satellites like Terra \citep{STEVENS2004405} and Aqua \citep{Parkinson2003Aqua} with wide-swath, low-resolution scanning capabilities were analyzed to look for events such as hot-spots created by volcanic activity. Based on other contextual information and pre-defined, site-specific thresholds, such detections triggered observation requests for a satellite with higher resolution imaging capabilities, like NASA's Earth Observing One (EO-1) \citep{Ungar2003EO1}. The sensor web also assigned a priority to each observation request. An automated mission planning system processed the requests, and carried out the observation if a) it could be fit within the schedule without disrupting higher priority observations and b) it did not violate any mission constraints. This sequence of operations is illustrated in Figure \ref{fig.ChienSensorWeb}.

The sensor web architecture was designed to ingest data from multiple sources, including ground-based sensor networks, terrestrial observatories and Earth-observing satellites. The concept was also adapted for the development of a flood sensor web \citep{chien_jais2019_using}, a wildfire sensor web \citep{chien-doubleday-mclaren-et-al-IGRSS-2011}, and a cryosphere sensor web \citep{doubleday-mclaren-chien-et-al-IJCAI-2011} for detecting events related to the break up of large polar ice sheets. %

\subsection{NASA Mars and Venus Autonomy Design Reference Missions}

\begin{figure}

	\centering
	
	\includegraphics[width=0.45\textwidth]{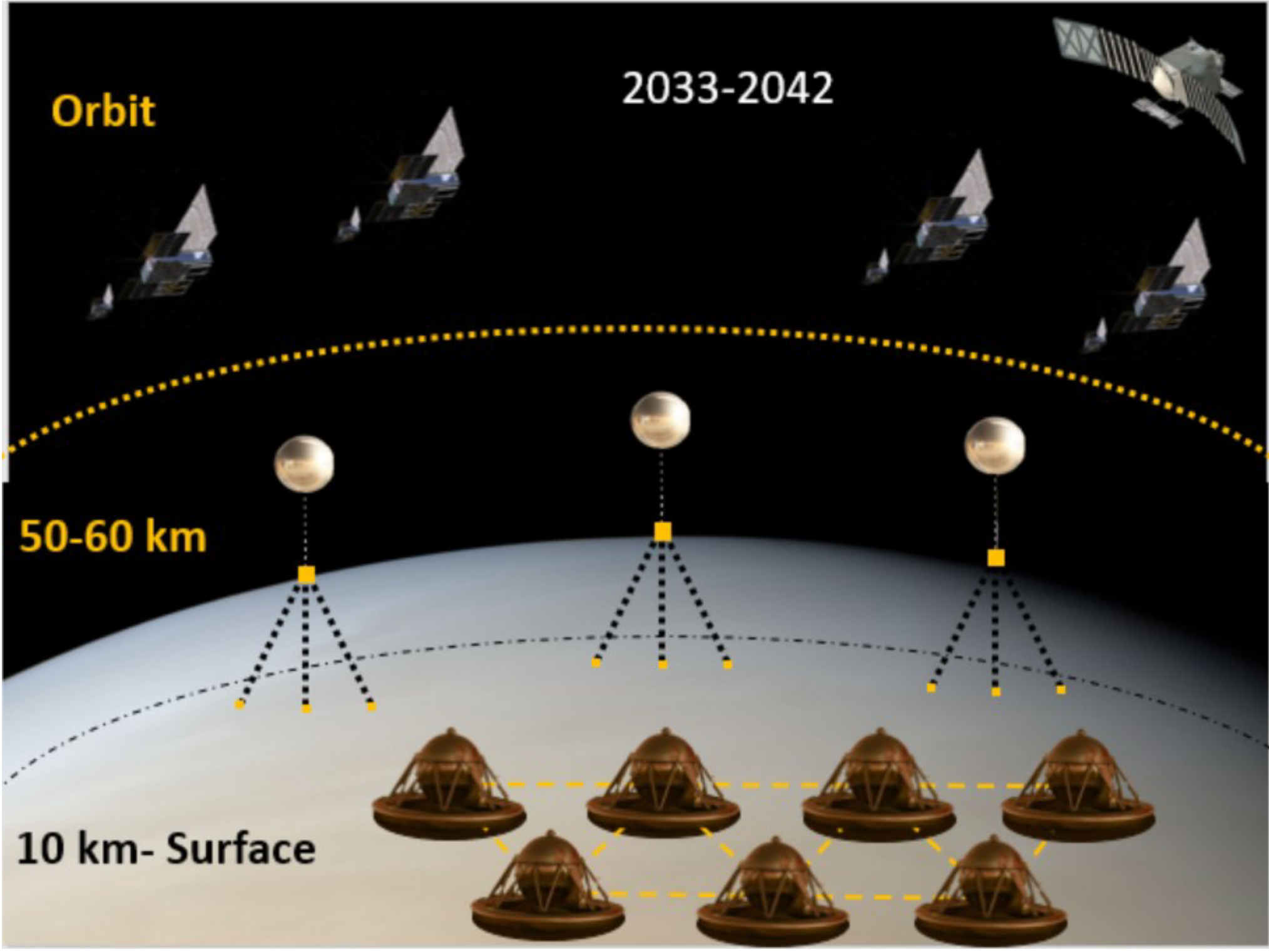}	
	\includegraphics[width=0.45\textwidth]{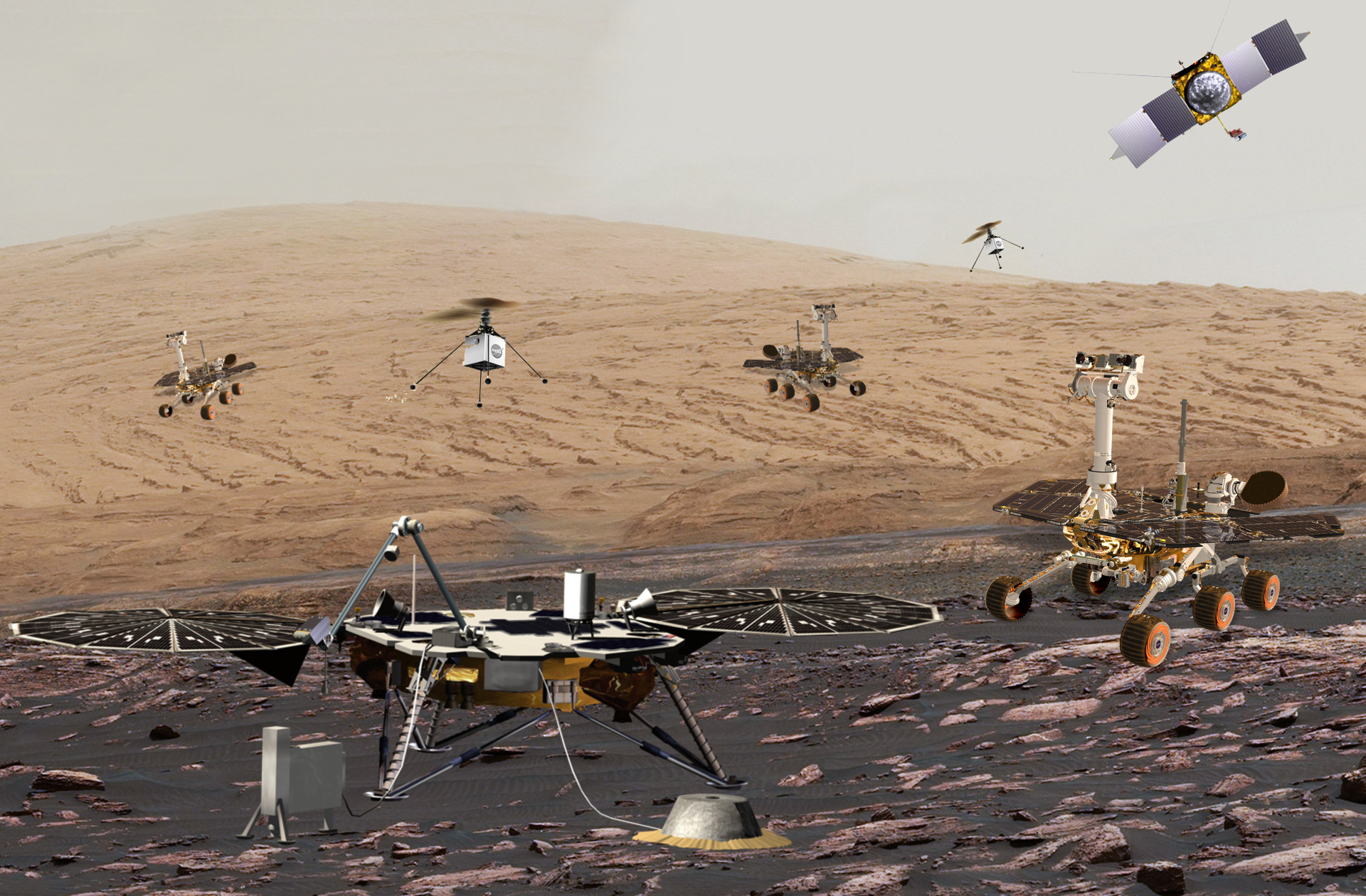}	
	\caption{Proposed design reference missions for exploration of Venus (left) and Mars (right) \citep{NASAAutonomyDRM2018, lyness2019marsdrm}.}
	\label{fig:history:mars-venus-drm}	
\end{figure}

In 2018, a NASA autonomy workshop drafted a set of reference design missions for Mars and Venus exploration to meet 2013-2022 Decadal Survey science objectives \citep{Decadal2011} and to identify key enabling technologies \citep{NASAAutonomyDRM2018, lyness2019marsdrm}.

The reference mission for Venus used a combination of landers, aerial platforms (e.g.,  balloons), orbiters, and atmospheric probes (Figure \ref{fig:history:mars-venus-drm}, left). Multi-agent coordination, with one platform building upon the observations of the other in an autonomous manner, was identified as a key technology enabler for this mission concept. 

Similarly, the reference mission architecture for Mars involved multiple assets including a fleet of small rovers (for sample acquisition and delivery to a lander); helicopters for atmospheric measurements and aerial imagery; one or more landers for detailed sample analysis with powerful computing and communication infrastructure, and one or more orbiters for Earth communications and global imagery.

Rovers would traverse outward from the landing site in a cooperative search pattern, collecting samples at intervals, helping scientists on Earth build a better picture of subsurface geohydrology. Collaborative, autonomous multi-agent task planning, response to scientific observations, and response to failures and anomalous scenarios were listed as key technology areas for future research. Efficient use of multi-agent resources would allow the system to function as a true distributed instrument. 

\subsection{Lunar Geophysical Network}

The 2013-2022 and 2023-2032 Planetary Science Decadal Surveys \citep{Decadal2011, Decadal2022} both identified as a high priority a Lunar Geophysical Network (LGN) mission to study the interior composition and structure of the Moon with a network of landers equipped with seismometers, magnetometers, heat flow sensors, and retroreflectors, operating through an entire lunar tidal cycle of six Earth years. A LGN mission concept was developed for the 2020 NASA Planetary Mission Concept Study Report, in preparation for the 2023-2032 Decadal Survey \citep{LGN2020}. 

The scientific goals of the mission concept include understanding the internal composition of the Moon, determining the distribution and origin of lunar seismic activity, and studying the thermal evolution, bulk composition, and magnetic field of our natural satellite.
All enabling technologies required for the LGN mission concept are at TRL 6 or above \citep[Appendix B]{LGN2020}. Due to the proximity to Earth, data compression is a comparatively minor concern.

\begin{figure}[h]
	\centering
	\ifaaspsj
	\gridline{
	\fig{lgn_2020_overview.pdf}  {.45\textwidth}{(a) Mission overview \citep[Figure 7]{LGN2020}.}
	\fig{lgn_2020_instruments.pdf}{0.45\textwidth}{(b) Accommodations for science instruments \citep[Figure 3]{LGN2020}.}
	}
	\else
	
	\begin{subfigure}[c]{0.4\textwidth}
	\includegraphics[width=\textwidth]{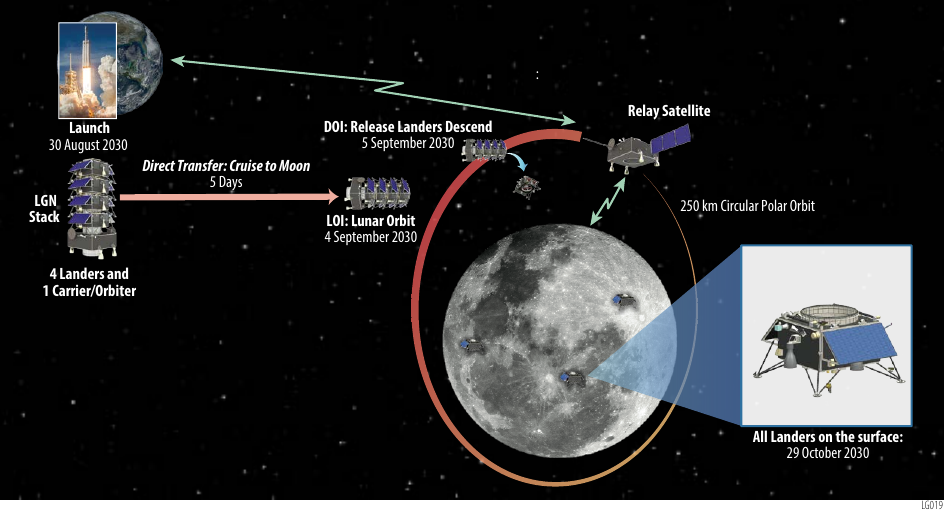}		
	\caption{Mission overview \citep[Figure 7]{LGN2020}.}
	\label{fig:history:lgn:mission}
	\end{subfigure}	
	\begin{subfigure}[c]{0.4\textwidth}
	\includegraphics[width=\textwidth]{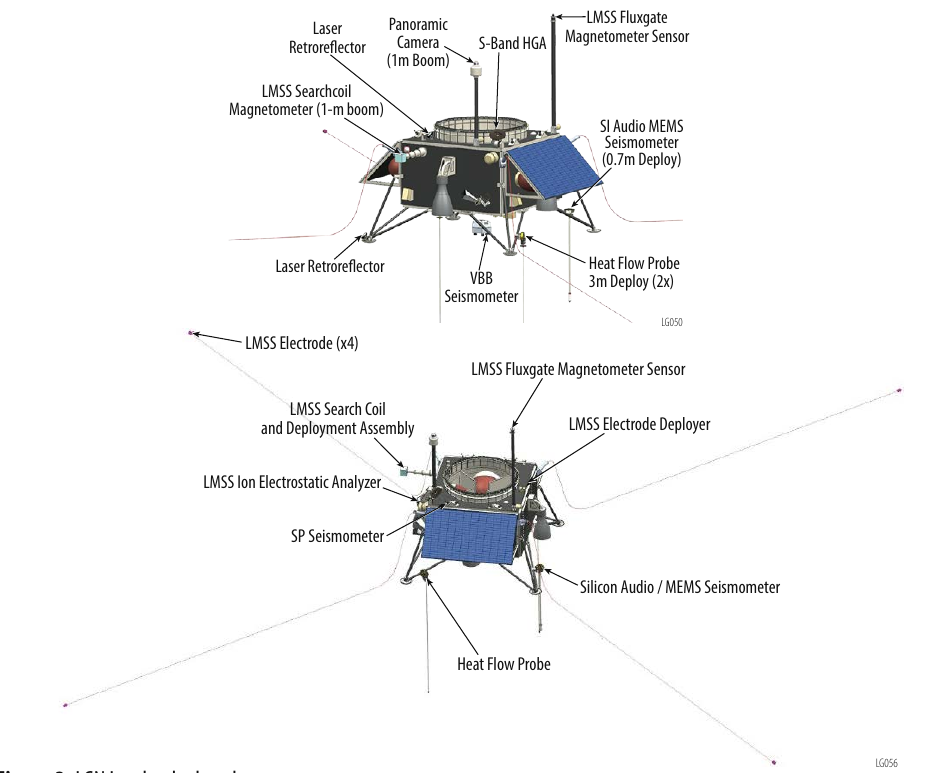}		
	\caption{Accommodations for science instruments \citep[Figure 3]{LGN2020}.}
	\label{fig:history:lgn:instruments}	
	\end{subfigure}
\fi
\caption{Lunar Geophysical Network mission overview and science instruments}
\label{fig:history:lgn}
\end{figure}

Figure \ref{fig:history:lgn} shows a high-level overview of the LGN mission concept and the science instruments carried by each LGN lander.

\subsection{Timeliness of Distributed Instruments}

NASA and the planetary science community have demonstrated a keen interest in distributed instruments, sensor networks, and multi-agent systems over the last thirty years, and this interest continues unabated, as shown by the recent NASA autonomy design reference mission (DRM) concepts for Mars and Venus exploration \citep{NASAAutonomyDRM2018, lyness2019marsdrm} and the Lunar Geophysical Network concept \citep{LGN2020}. 

Why is it that no distributed instrument has flown on a planetary exploration missions to date, then? And why could this time be different?

The rest of this paper aims at answering this question. Specifically, in Section \ref{ch:sciencecase}, we justify the ``science pull'' for distributed instruments by showing that such instruments are uniquely well-positioned to address crucial and timely questions in planetary science, in line with the findings of the 2018 NASA Autonomy DRM concepts \citep{NASAAutonomyDRM2018, lyness2019marsdrm} and with Decadal priorities. We also show that many high-priority science questions require regional or global networks, with distances of hundreds to thousands of kilometers between sensors; this partially explains why the SensorWeb project, which developed technologies for local sensor network, did not see adoption in flight.
Next, in Section \ref{ch:technologies}, we survey required technologies for distributed instruments. Recent technology advances have filled many of the gaps that hindered adoptions of distributed instruments in the past. In particular, compared to the 1990s MESUR architecture, advances in sensor miniaturization, communication hardware, and power solutions make smaller landers possible, greatly reducing the complexity and cost of planetary-scale sensor networks.
 
\section{The Science Case for Distributed Instruments}
\label{ch:sciencecase}

Distributed instruments hold unique promise for scientific questions that require the investigation of spatio-temporal correlations at regional and global scale.

To assess the promise of distributed instruments for a variety of scientific disciplines, we considered a range of planetary science questions, drawn from the scientific literature and the latest Planetary Science and Astrobiology Decadal Survey, and applied an objective set of criteria to evaluate which could most benefit from a feasible distributed instrument formulation.

\subsection{Methodology}
The following criteria were used in the search.

\begin{itemize}
\item \textbf{Compelling to a broad scientific community}: We chose to focus on high-impact science questions with proven interest from the scientific community; specifically, we turned our attention to  scientific topics explicitly identified in the two most recent Planetary Science Decadal Surveys \citep{Decadal2011,Decadal2022}.

\item \textbf{In-situ Planetary Science}: We chose to focus our attention on \emph{in-situ planetary science}, excluding remote sensing solutions such as spacecraft swarms, orbiting distributed apertures, and space-based interferometers.
This choice was motivated by a desire to bound the problem space, allowing for a more focused evaluation; we refer the interested reader to \cite{bss2018swarms} for an assessment of the potential of swarms of space vehicles for remote sensing and astrophysics.

\item \textbf{Uniquely Beneficial}: A focus was placed on science questions that can be \emph{uniquely} addressed by distributed instruments; that is, science questions that can only be satisfactorily answered through the collection of multiple geographically distributed, spatially and temporally correlated measurements. In the selection of scientific questions of interest, oft-touted benefits of distributed systems such as lower cost, higher resiliency, and faster data collection were not considered, unless such benefits had a direct impact on the ability of the proposed architecture to address the science question of interest. 

\item \textbf{Feasible}: When identifying scientific questions of interest, we performed a deliberately loose feasibility check, adopting an optimistic outlook on future technology developments and integration costs, to avoid overconstraining the search. Once science questions of interest were identified, a more detailed feasibility assessment was performed for the top contenders: this assessment is presented in Section \ref{sec:technology:gaps}.

\item \textbf{Failure is an option} The goal of the search was explicitly \emph{not} to directly find applications for distributed systems. Specifically, it was accepted that there might not exist any \emph{compelling} planetary science questions that could be \emph{uniquely} addressed by \emph{feasible} distributed instruments (e.g., because such questions can be addressed satisfactorily by monolithic instruments). By adopting this perspective, it was ensured that the outcome of the search, if any, would be highly compelling from a scientific perspective, and would justify the additional complexity introduced by distributed instruments.
\end{itemize}

Based on these criteria, we identified four broad areas of scientific inquiry that satisfied the aforementioned criteria: global climate and weather, localization of seismic events, magnetometry for internal composition, and trace gas detection and localization. Each of these areas is discussed in detail next.

\subsection{Global Climate and Weather}
\label{sec:sciencecase:climate}

Understanding atmospheric circulation is a key scientific priority for rocky bodies with an atmosphere, e.g., Venus, Mars, and Titan. The investigation of Martian atmospheric phenomena is particularly compelling due to its relevance not only to fundamental questions of planetary formation and evolution, but also to EDL and, eventually, to human exploration.

Mars climatology scientific priorities align along three themes, corresponding to MEPAG Objective II.A, to ``...characterize the state and controlling processes of the present-day climate of Mars under the current orbital configuration'' \citep{BanfieldEtAl2020MEPAG}:
\begin{itemize}
\item Surface-atmosphere interface, including exchange of material (e.g., moisture and methane), and transport phenomena on the surface (e.g., dust);
\item Global atmospheric circulation;
\item Transient events in atmospheric circulation, e.g., atmospheric tidal behaviors and the evolution of dust storms.
\end{itemize}

The key challenge is that past investigations provide \emph{global} records from remote sensing, with extremely low temporal resolution, and \emph{hyperlocal} measurements from rovers and landers.

To fill this gap, one can either deploy a fleet of orbiters with phased overflight times, or a number of ``embarassingly parallel'' landers to provide highly correlated measurements.
While macroscopic and planet-level phenomena have seen significant investigation, a number of key scientific questions (e.g., the seasonal transition of the Hadley circulation and its influence on, and response to, atmospheric dust,
the tidal behavior of the Martian atmosphere at the diurnal and semi-diurnal scale, and the growth of dust storms) are not satisfactorily addressed, and local phenomena at the microscopic scale (which are strongly influenced by topography, surface composition, etc.) are largely unknown. Distributed instruments hold promise to fill this gap.

\subsubsection{Science and strength of distributed instrument}

Several decades of spacecraft study of the Martian atmosphere have revealed a complex global climate system that is, in many ways, more complex than Earth’s own.  Despite the absence of liquid water or biosphere, the Martian surface environment is a complex interplay of several interrelated climate cycles.  Even with ample prior effort invested in quantifying atmospheric circulation, composition and interaction with the surface, much remains poorly understood about the Martian atmosphere, reflecting several cost constraints and technological weaknesses in these prior efforts.  In addition to the value of understanding the Martian atmosphere from a scientific perspective (Mars, for example, provides a comparative end member among terrestrial planetary climates in our Solar System), proper characterization of the global environment is essential for reducing risk to future spacecraft missions to the Red Planet.  Entry, descent and landing on the Martian surface relies on knowledge of atmospheric density and wind profiles near the proposed landing site.  Future efforts to enable even higher-precision landing will require a correspondingly better understanding of atmospheric behavior.  Direct spacecraft observations (from orbit or the ground) provide only limited coverage, both temporal and spatial, of the Martian atmosphere.  Filling in these gaps requires the use of numerical models that can simulate atmospheric behavior at all locations and times of day, but which are constrained by the limited availability of direct observations with which to validate the models.  Specifically, these general circulation models rely on two orthogonal types of data sets to ensure they are able to simulate the Martian atmosphere in a reasonable fashion.  Landed spacecraft generally provide high temporal resolution of atmospheric behavior, but only at a single site, so they have, inherently, very poor spatial coverage.  Conversely, orbiting spacecraft can provide near-global coverage, but with limited temporal resolution (a typical polar-orbiting satellite in a Sun-synchronous orbit has only two overflights per day, each 12 hours apart).  Together, these data sets provide a piecemeal overview of Martian planetary climate.%

Some key outstanding questions about the Martian atmosphere are best addressed by more sophisticated observational approaches, such as joint surface-orbit measurements, or simultaneous observations at multiple locations across the planet.  As an example of this growing need, \emph{characterization of the dynamical and thermal state of the lower atmosphere} is identified as a higher priority investigation in the MEPAG Goals document \citep[Goal II, Investigation A1.1]{BanfieldEtAl2020MEPAG}.  The higher priority nature of this investigation is meant to indicate that existing measurements are insufficient to address the investigation, and are necessary to achieve the overarching MEPAG objective (\emph{“Characterize the…present-day climate of Mars”}).  Under this investigation is mentioned, specifically, the need to characterize larger-scale circulations such as the thermal tides.  The behavior of the Martian thermal tides—global-scale oscillations in atmospheric properties—makes its study ideally suited for measurement by an instrument with wide global coverage.

\begin{figure}[h]
\centering
\includegraphics[width=.8\textwidth]{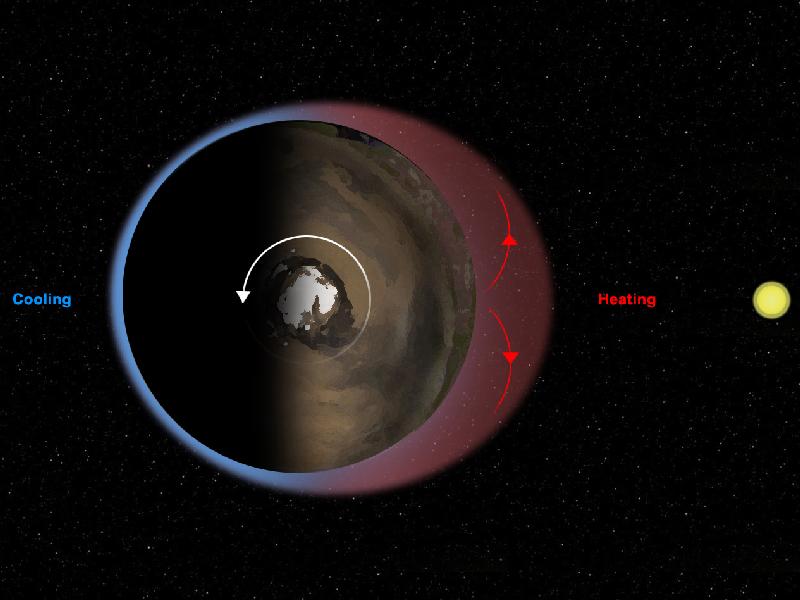}
\caption{Illustration of Mars’ thermal tide, responsible for large, daily variations in pressure at the Martian surface.  Heating by the Sun drives global circulation of air towards the night side of the planet, resulting in a diurnal cycle of surface pressure as the planet rotates beneath the Sun.  The timing and structure of the thermal tide is dependent upon such elements as atmospheric dust and topography.  Image credit: NASA/JPL-Caltech/Ashima Research/SwRI.}
\label{fig:pathfinders:thermal-tide}
\end{figure}

The strong atmospheric tides in the Martian atmosphere (Figure \ref{fig:pathfinders:thermal-tide}) can yield daily changes of surface pressure of as much as 10\% at a single location \citep{haberle2014preliminary}.  These tides, which vary on diurnal and sub-diurnal timescales, are strongly impacted by topography \citep{withers2003effects} and atmospheric dustiness \citep{wu2020dust}, yet we have only a limited grasp of how they function on Mars, based largely on poor spatial and temporal coverage.  Orbiting spacecraft capture the same long-period phases of the tide (the diurnal and semi-diurnal components), but provide little information about variability of the cycle over the course of a single day.  Landed spacecraft can capture elements of the higher frequency tidal modes, but, absent multiple sensors on a global scale, they are not able to interpret the full range of motion that makes up the tidal oscillations.

\begin{figure}[h]
\centering
\includegraphics[width=.8\textwidth]{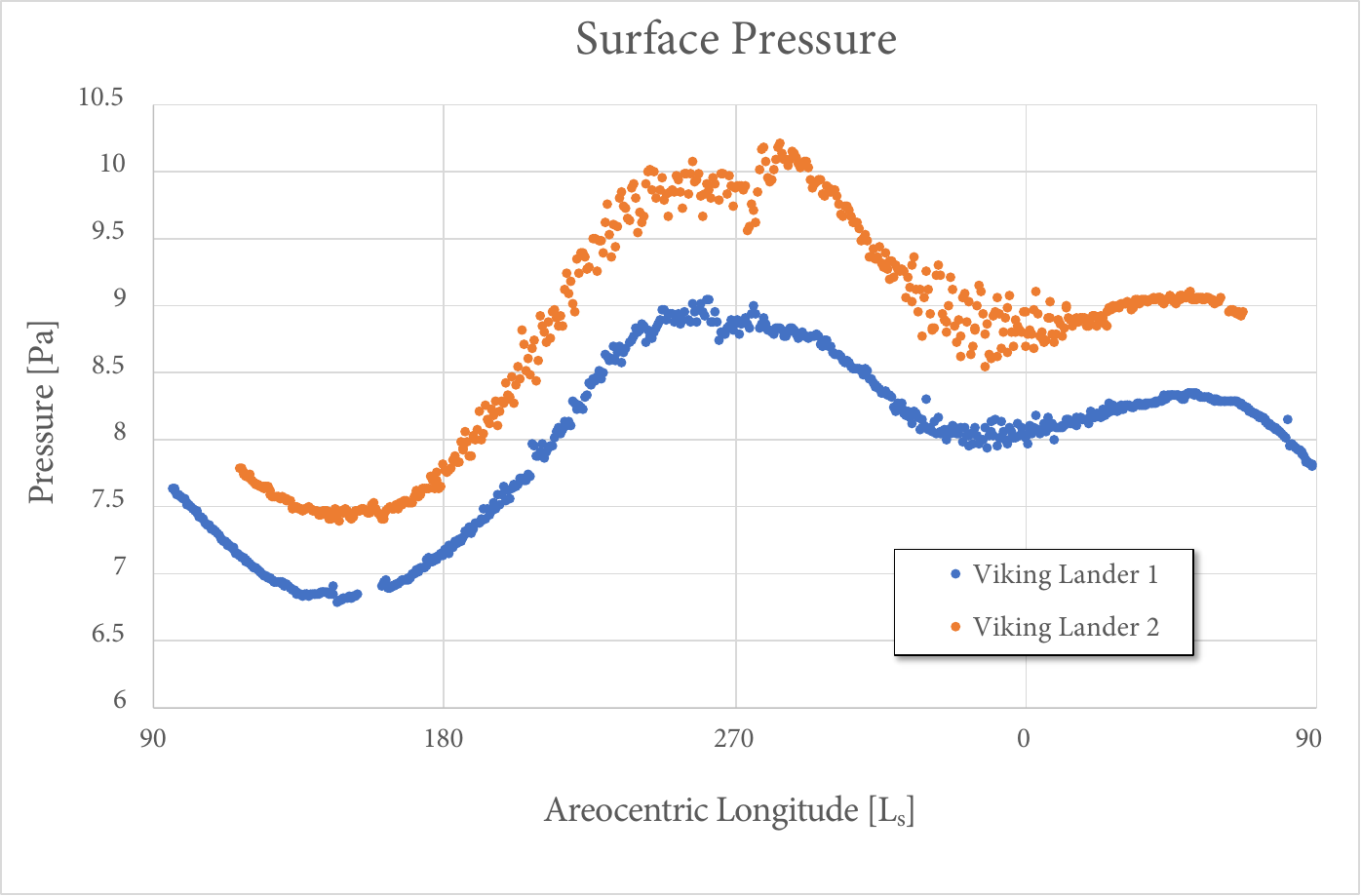}
\caption{The annual pressure cycle as measured by Viking Landers 1 and 2 shows a ~30\% variation between peak and trough.  Offset of the two curves is due to different elevation of each lander.}
\label{fig:pathfinders:pressure-cycle}
\end{figure}

A suite of integrated sensors across a wide range of surface locations would enable a proper visualization of the global motions of the atmosphere, as the atmosphere is not a series of independent, isolated locations but, rather, a globally integrated unit. Because of the global nature of these desired investigations, they require instruments that are global in scale—for this reason, we envision a distribution of multiple sensors as necessary to address these questions.  Rather than a suite of independent measurements, this \emph{distributed} instrument should be seen as a single, global scale measurement obtained by multiple sensors operating simultaneously.  This perspective is what distinguishes the present approach from prior efforts for obtaining global coverage.
Similarly, the global condensation flow, a consequence of the annual CO$_2$ condensation/sublimation cycle on Mars, may be best observed through a suite of integrated sensors spanning the planet.  Figure \ref{fig:pathfinders:pressure-cycle} illustrates the seasonal cycle of pressure as measured at the Viking Lander 1 and 2 sites.  As ~95\% of the Martian atmosphere is CO$_2$, surface pressure is a good proxy for total atmospheric CO$_2$ abundance.  The large seasonal variation of about 30\% is a consequence of CO$_2$ condensation in the cold wintertime hemisphere, which pulls mass out of the atmosphere and onto the polar surface, followed by sublimation of these polar ice deposits back into the atmosphere during spring and summer.  A careful look at Figure \ref{fig:pathfinders:pressure-cycle} reveals asynchronous timing between the local maxima/minima between the two curves.  This is because the sublimated mass of CO$_2$ released from the polar cap does not instantaneously ‘flood’ the atmosphere but is, rather, distributed more slowly and in a non-uniform manner, affecting the two landing sites differently, and at different times (high frequency ``spikes'' are due to weather patterns, and are distinct from the seasonal behavior of the condensation cycle).  As the spring/summer season progresses, the location of these sublimating deposits changes \citep{piqueux2015variability}, as does the nature of this ``condensation flow''.  By capturing the evolution of the pressure cycle at multiple locations simultaneously using a distributed instrument, we can better understand the timing of condensation/sublimation activity and the polar cap composition as a function of season.  This, too, is identified as a higher priority measurement in Mars science \citep[Goal II, Investigation A2.2]{BanfieldEtAl2020MEPAG}.  As noted in \cite{BanfieldEtAl2020MEPAG}, “Due to poor local time coverage in existing observations (a result of sun-synchronous spacecraft orbits), existing observations have not been able to measure this [CO$_2$ sublimation] variability.”  A globally distributed instrument could directly address this need.

\subsubsection{Distributed Instrument Configuration and Strength}

\begin{figure}[h]
\centering
\includegraphics[width=.5\textwidth]{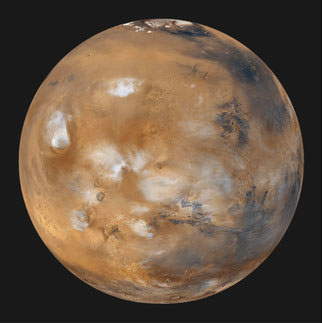}
\caption{A global image of Mars showing the presence of water ice clouds in the lower latitudes and near large topographic features.  This collection of clouds is referred to as the ‘aphelion cloud belt’ and occurs primarily during the late northern spring/early summer, when Mars is furthest from the Sun.  Image credit:  NASA/JPL-Caltech-MSSS}
\label{fig:pathfinders:ice-clouds}

\end{figure}

\begin{figure}[h]
\centering
\includegraphics[width=.8\textwidth]{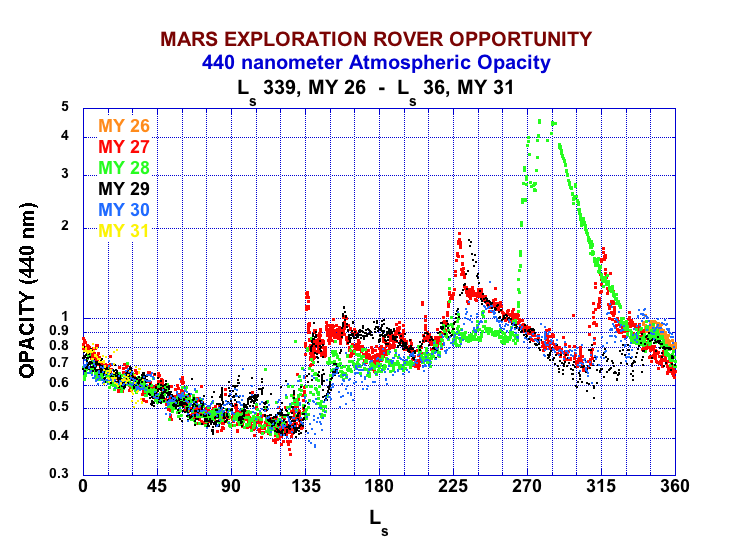}
\caption{Six years of atmospheric opacity data illustrating the seasonality of atmospheric ‘dustiness’ on Mars.  Opacity is noticeably higher in the second half of the year, which corresponds to southern spring and summer, when Mars is closest to the Sun.  Data obtained by the Mars Exploration Rover ‘Opportunity’, from the PDS Planetary Atmospheres Node at \url{https://atmos.nmsu.edu/data_and_services/atmospheres_data/MARS/MER/spirit/atm_opacity.html}}
\label{fig:pathfinders:opacity}
\end{figure}

There are three cycles that drive climate on Mars—the aforementioned carbon dioxide cycle, the water cycle (both as vapor and ice), and the dust cycle.  Atop the main CO$_2$ cycle are seasonal variations in water vapor and water ice abundances in the atmosphere, both of which peak, generally, during the northern spring and summer seasons.  A widespread belt of clouds, for example, forms in the tropical latitudes during late spring/early summer [\cite{clancy1996water}; Figure \ref{fig:pathfinders:ice-clouds}]. During southern spring and summer, however, the Martian atmosphere is generally dustier than at other times of the year (Figure \ref{fig:pathfinders:opacity}), though dust is ubiquitous in the Martian atmosphere.  There is a strong relationship between atmospheric dust and water ice, as the former serves as condensation nuclei for the latter; hence, one cannot observe one cycle without considering the other.  Because of both the temporal and spatial distinction between the `‘key’' areas of the different cycles, observations need to be made across a wide spatial extent, and across multiple seasons, if not years.  Global coverage of Mars with identical sensor packages as part of a distributed instrument would therefore be a valuable approach to providing this global coverage and achieving good temporal resolution.  A strawman design for a distributed instrument might have on the order of 10 distinct sensor packages roughly covering all longitudes and latitudes (4 latitudes by 4 longitudes would provide quadrant-scale coverage and result in 16 sensors).

A threshold sensor package that would positively advance Mars atmospheric science would contain pressure and air temperature sensors, as well as sensors to measure wind, and column-integrated dust and water ice abundance.  Augmentations to this basic package might also include a filtered imager or spectrometer to measure clouds and atmospheric opacity, as well as water vapor abundance.  Sensors probing the surface could enable additional useful measurements of surface-atmosphere fluxes of water and temperature, and an understanding of soil composition.  The basic package is largely comprised of high-TRL, low-cost instruments (pressure, temperature and wind sensors have all successfully flown to Mars, e.g., \cite{gomez2012rems,spiga2018atmospheric}), which is an enabling factor, in terms of cost and risk, for deploying multiple sensors across the planet.  Adding additional components to the sensor package increases its overall scientific value, but at the price of greater expense or increases in SWaP and with lower technological readiness.  Examples of such sensors are a Doppler lidar (for cloud/aerosol detection and wind; e.g., \citep{whiteway2008lidar}) or sonic anemometer (for high-frequency wind measurements; e.g., \cite{banfield2016martian}).  While this study recognizes the potential value of such measurements, they may be of secondary value towards addressing global-scale atmospheric processes (versus local atmospheric behavior), for which a distributed instrument is ideally suited.

\subsubsection{Technical Challenges}

The proposed sensor package has a high TRL, greatly reducing the technical hurdles for development of a distributed instrument for global Martian climate and weather. Two key technical challenges of such an instrument, however, are cost-effective delivery to the planetary surface from orbit, and long-term survivability -- i.e., the need to deploy multiple sensors across a planetary surface with lower per-sensor costs compared to current monolithic missions, and the requirement to ensure the sensors are able to collect data for months to years across a variety of landing sites, including especially inhospitable polar regions. 
\revmm{%
These challenges are broadly common to \emph{all} instrument concepts discussed in this section. For this reason, we defer their detailed discussion to Section \ref{sec:sciencecase:common-challenges}. %
}

\subsection{Seismic Techniques: localization of seismic events}

Localization of seismic events
is critical to illuminate active geological processes \revmm{on rocky and icy bodies}, giving insight into planetary evolution. While localization can be performed with a single instrument by carefully measuring arrival times and polarization, the process is highly challenging and requires extremely high-quality data: for instance, data from  \revmm{InSight's SEIS instrument} (which is equipped with a \mmmargin{\revmm{very-broad-band seismometer co-located with a short-period seismometer}}{Not to be pedantic, but technically the SEIS 'instrument' is composed of two seismometers--the VBB seismometer and the SP seismometer}) has been used to detect 2769 potential seismic events, but only 42 of these events have been reliably localized \citep{InSightSeismicCataloguev132023}.
In contrast, having multiple geographically separated seismometers holds promise to make localization of seismic events \emph{many orders of magnitude} easier, drastically increasing the number of events that can be detected by sensor packages in a given class, and reliably providing insight into active geological phenomena.

\subsubsection{Science Benefits of Distributed Seismology}
Planetary surface seismology is one of the critical techniques for investigating the interior of various objects in the Solar System, and distributed seismology stands out as one of the prime examples of science that is enabled by  distributed instruments. 
The field of seismology lends itself naturally to distributed measurements, since such measurements readily allow for the localization of the source of seismic activity, which is critical to answering questions about the interior structure of the body. %
The 2023-2032 Decadal Survey \citep{Decadal2022} explicitly calls for geophysical networks capturing seismic and magnetometer measurements on the Moon and Mars (Q3.3, Q5.1), and for seismic investigations (although not explicitly seismic networks) of Venus (Q3.4) and of rocky and icy worlds (Q5.6), with particular focus on Jupiter and Saturn satellites including Europa, Enceladus, and Titan (Q8.2).

Seismic waves can be generated from tectonic and fluid flow processes, and their observations can not only tell us about the internal structure of the celestial body, but also shed light on internal dynamic processes. Ocean worlds with significant tidal deformations and bodies with ongoing seismic activity (e.g., plumes), such as Enceladus and Titan, are prime candidates for seismic exploration. Seismology can answer key questions about the thickness of Europa's icy crust, localize the geothermal activity responsible for the plumes emanating from the vents on Enceladus, and study the intensity of active volcanoes on Venus. Some of the specific planetary science questions that can be explored with the help of distributed seismic instruments are:

\begin{itemize}
	\item 3D seismic structure of any solid-body object in the Solar System;
	\item Localization of seismic activity associated with plume processes on ocean worlds;
	\item Measurement of the distribution of magma on Io;
	\item Detection of ice subduction on Europa or other ocean worlds;
	\item Measurement of surface tilt caused by tidal flexing;
	\item Flooding of a region of interest with a large number of seismometers (a few hundred) to obtain a very dense map for active surveying, akin to how targeted regions are explored in the oil industry.
\end{itemize}

\subsubsection{Configuration of the Distributed Instrument}
At least 3 seismometers (4 preferable) are needed to enable triangulation-based localization of seismic events. In terms of spatial distribution, if there is poor a priori knowledge of active regions, then distribution on a global scale would be the preferred option with the aim of discovery and exploration. A global distribution covering a wide range of longitudes and latitudes is also favorable for increasing localization accuracy. On the other hand, if there is good knowledge of active seismic locations from orbiter missions or other sources, then a tighter configuration might be more favorable. The efficacy of a regional network with moderate separation between individual sensors is yet to be analyzed in detail. \cite{Mocquet} studied a global Mars network comprising four hubs, each of which contained a mini-network of four distributed seismometers. Global coverage using a single launch is hard to achieve for objects like Mars due to guidance and EDL challenges. However, a global seismic network becomes more feasible on smaller bodies like Enceladus, where each sensor node has to be imparted with relatively small amounts of $\Delta$V, and survive lower impact speeds. %

The performance specifications for each individual sensor, such as sensitivity, noise figure, bandwidth, etc. have to be tuned for each target body. Recent advances have led to the development of miniaturized short-period seismometers such as the SEIS instrument flown on the Mars InSight mission \citep{SEIS}. Other such instruments are also currently in development for the Lunar Geophysical Network \citep{Shearer2011LGN,Neal2019LGN}.

\subsubsection{Unique Technical Challenges}

\paragraph{Data handling and data transfer}
A major technical challenge posed by distributed seismology is satisfying their demanding data handling and data transfer requirements. Seismometers are looking for sporadic events and, hence, require near-continuous operation to increase the likelihood of detection. This typically translates to data rate requirements on the order of a few kHz. On-board computation and analysis of the waveforms and automated detection of scientifically relevant events can reduce the amount of data that needs to be relayed back to the Earth.
A dedicated  effort is needed to develop algorithms for on-board data processing to reduce the volume of data that needs to be downlinked. These algorithms would have to be parameterized in a way so that they can be easily tuned upon deployment and adjusted based on the signals encountered during the mission. In the past, it has been challenging to automate this process, since poor priors are available on the expected structure and frequency of signals. Lightweight algorithms that can be deployed on low-power embedded computing systems can help overcome the critical challenge of data reduction. 

\paragraph{Calibration, localization, and synchronization} The use of tri-axial seismometers reduces the sensitivity of the instrument to the orientation of the sensor on the surface; however, the sensors have to be localized with a reasonable accuracy, and the clocks on the sensors have to be tightly synchronized, to enable the localization of seismic events. Other challenges involve the need for calibration against noise sources such as wind and thermal fluctuations, and the potential need for anchoring to the planetary surface.  For this reason, it worthwhile to explore the possibility of adding a meteorological package alongside the seismometer to eliminate the effects of temperature, wind, etc., on the seismic measurements.

\subsection{Magnetometry for Internal Composition}

Magnetometry is a powerful tool that can be used to probe the internal structure of planetary bodies and gain insight into the evolution of a past (or current) dynamo. Geophysical magnetometer investigations can be applied in three ways: electromagnetic sounding, sensing an active dynamo, and sensing remanent crustal magnetization. Active dynamos and their associated magnetospheres can be sampled satisfactorily by one or more orbiters (such as the MMS mission \citep{burch2016magnetospheric}), and do not require distributed networks on the surface. However, the application of distributed instruments would be highly useful to sound the interiors of solid bodies and for future planetary exploration of remanent crustal magnetization.

Electromagnetic sounding can be used to understand the present-day internal layering within a planetary body. Electromagnetic sounding is a technique which uses magnetometry to sense the conductive layers within the body, such as an iron core or a salty ocean. A conductor in the presence of a time-varying magnetic field will produce eddy currents, which create an induced magnetic field that opposes the time variation of the primary field. This secondary field can be detected by identifying perturbations to the primary field. In planetary environments, a time-varying magnetic field may be supplied by the interplanetary magnetic field that originates from the sun or by the relative motion of a planet’s magnetic field to its moon \citep{russell1981measurements,khurana1998induced} . As an example, for an icy moon like Europa, Jupiter’s rotating magnetic field acts as the primary time-varying field, and the measurement of interest would be the perturbations to this field caused by magnetic induction via Europa’s ocean. The depth to which the interior can be sounded depends on the frequency of the inducing magnetic field and the electrical conductivity of the various layers within the body.

In addition, the presence of a past dynamo can be inferred by detecting remanent magnetization in the surface crust. This occurs if its superficial composition contains ferromagnetic grains that align with the magnetic field direction as the crust cools. Hence, crust formed during different time periods can contain snapshots of its contemporaneous magnetic field. This magnetization can be erased via impact heating or chemical alteration, which may leave the surface with a “patchy” distribution of surficial magnetization (as has been seen on the Moon and Mars). Measurements of crustal magnetic fields allow constraints to be placed on the strength and direction of the field at the time of formation. Further, if the age of the magnetized crust can be dated, the lifetime of the planetary dynamo and the evolution of its internal conditions may be deduced. Understanding the history of a body’s dynamo evolution is important not only for determining conditions early in its interior, but also for assessing its past habitability \citep{Brain2020Magnetic}. Intrinsic magnetic fields may affect atmospheric evolution through their effects on atmospheric ion loss rates due to solar wind interaction, which are believed to vary with the body’s magnetic field strength \citep{egan2019planetary,sakata2020effects}. 

Remanent crustal magnetization has previously been detected at the Moon and Mars from orbit using instrument packages that combined magnetometers and electron reflectometers to detect electrons magnetically reflecting from crustal magnetic fields to infer the surface magnitude \citep{lin1979constraints,acuna1999global}. However, the spatial resolution of this technique is limited by the gyroradius of the electrons in the planetary environment; in the lunar case, the resolution limit is ~10 km \citep{lin1979constraints}. Thus, to obtain high resolution crustal maps of a planetary surface would require surface measurements with a magnetometer.

\subsubsection{Science Benefits of a Distributed Magnetometry Instrument}

Magnetic induction of global electrically conductive layers results in a dipole field whose signal falls off as $1/R^3$. Therefore, it is ideal to be as close to the surface of the body as possible to measure the time-varying induced field, especially since some responses may have low amplitude. In addition, surface magnetometers provide continuous coverage, while orbiters may only be able to provide informative data when the spacecraft is within a certain altitude range (i.e., near periapsis). Since inversion methods require enough magnetic data to properly sample the body’s response to the variation of the inducing field, continuous coverage from the surface can lead to a more robust solution for determination of interior layers. A network of strategically placed magnetometers carrying out simultaneous measurements would be needed to build a global view of the body’s time-varying induced magnetic field. 

\begin{figure}[h]
\centering
\includegraphics[width=.6\textwidth]{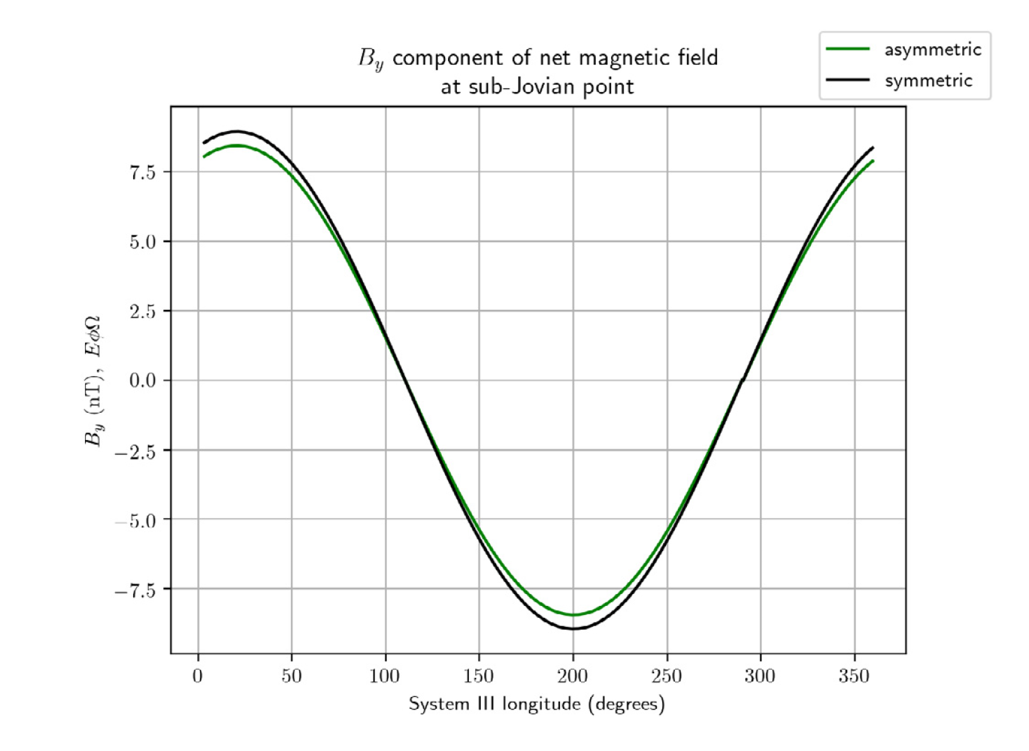}
\caption{ Differences in the $B_y$ component of the induced magnetic field of Europa in the $EPhiO$ coordinate system for a spherically symmetric ice shell (black) and an asymmetric ice shell (green). The difference in response only differs by a few $nT$ as seen from the subjovian point on the surface. Figure from \cite{styczinski354induced}.}
\label{fig:pathfinders:magnetometry:europa}
\end{figure}

It may also be possible for surface magnetometers to detect asymmetry in ice shells of ocean worlds. Recent literature has examined the differences in the induction response between a spherical ice shell and an asymmetrical ice shell \citep{styczinski354induced}. Figure \ref{fig:pathfinders:magnetometry:europa} shows variability in response for the case of Europa with a spherical (black line) and asymmetric (green line) ice shell for a surface location at the subjovian point. It can be seen that these differences are quite small and would require good sensitivity even at the surface location. Since the signal falls as $1/R^3$, it is doubtful that these differences would be measurable from orbit. A distributed network could possibly measure this by placing three surface magnetometers along each axis of a chosen coordinate system relative to the body’s center. Another prime target for such an investigation would be Enceladus, which has plumes located only at its southern pole. The current line of thought for this plume asymmetry is that the ice shell may be thinner near the south pole and thicker near the north pole. However, more analysis is required to understand how feasible an asymmetrical detection would be, and which planetary bodies may present an asymmetrical ice shell that could be detected through magnetometry.

A distributed magnetometry instrument would also provide additional scientific measurements. In addition to the induced magnetic field, the surface magnetometers would accurately sample both the spatial distribution and magnitude of crustal fields. While temporally coordinated measurements are not needed for mapping these time-invariant crustal fields, the multi-point measurements from geographically distributed sensors allows quick development of a more complete picture of the distribution of the remanent magnetic field on the planetary surface. 

\subsubsection{Distributed Instrument Configuration}

Magnetometers require modest power ($\sim 0.8$ W) and data rates ($\sim 30$ bps) \citep{Villareal2020decadal}, making them an instrument suitable for a distributed network. The optimum configuration of a distributed network on the surface will vary with the target’s environmental conditions and the type of magnetometry investigation. Crustal magnetism could be sensed by stationary, strategically placed  magnetometers (similar to InSight's) to compare different magnetized locations on the surface; this would allow for monitoring of how crustal fields interact with the solar wind and the ionosphere over time. However, networks of mobile platforms would likely be most ideal to sample and map a larger spatial range of crustal fields. A generic case for electromagnetic sounding could possibly consist of one spacecraft in orbit to directly measure the inducing field, and a small number (3 minimum) of magnetometers on the surface for greater spatial resolution at higher precision. Placement of the magnetometers on the surface may require a global distribution (such as near the poles and equator) to capture the induction response. However, detailed analysis is needed to identify preferable placements for networks on specific bodies. 

\subsubsection{Unique Technical Challenges}

\paragraph{Advance knowledge and potential precursor missions} In both applications discussed, prior knowledge of the magnetic environment would be required before a distributed network can be deployed. This is because the necessary sensitivity and the range of field strengths that the magnetometer will encounter need to be known. Thus, this requires that a preliminary exploration of the body be performed by an orbiter. For the case of magnetic induction, a companion orbiting spacecraft would also be needed in addition to a surface network to record the primary field. An orbiting spacecraft would also be useful (though not necessarily required) for the study of crustal fields, to place any temporal variation into context.

\paragraph{Calibration} Technical issues of such a network need to also be considered. For example, magnetometer measurements will be affected by the ambient temperature. Consequently, landing locations that may experience large diurnal temperature fluctuations (particularly at airless bodies) may need a mechanism for regulating the temperature of the instrument. Magnetometers are also typically placed on booms to avoid contamination from magnetic fields originating from the payload itself. Additionally, intercalibration of the magnetometers will be needed to allow for direct comparison of data at different locations.

\subsection{Trace Gas Detection}
\label{sec:sciencecase:tracegas}

Transient plumes of trace gases such as methane near the Martian surface are highly tantalizing, as they potentially hint at the presence of biological phenomena on other worlds. 
Identifying and localizing sources of gas detection, assessing emission fluxes, and measuring the methane/ethane ratio can shed light into the source of these plumes and determine whether they are caused by biological phenomena or by abiotic processes (e.g., serpentinization).
Orbiting assets can detect large gas releases, but they do not provide sufficient resolution to localize them, or to identify smaller plumes; individual surface assets, in turn, can serendipitously detect smaller gas releases, but they cannot easily identify or localize their origin.
In contrast, a network of gas sensors can greatly improve localization and detection of trace gases, providing simultaneous, spatially separated measurements with very high resolution; a distributed instrument can also measure gas fluxes, shedding further light on the nature of the phenomena and proving or disproving its biological nature.

\subsubsection{Science Benefit of a Distributed Instrument}

A high-priority scientific goal identified in the Mars Exploration Program Analysis Group Goals document \citep{BanfieldEtAl2020MEPAG} is to “determine if Mars ever supported, or still supports, life” through surface and subsurface smart sensor systems that can define, for example, the origin and point sources of methane and help determine whether Mars is still active (e.g., heat energy generated by magma and tectonism) and contains groundwater reservoirs, all important factors for life as we know it.

Life, hydrothermal activity (releases related to the mixing of magma and water), and serpentinization of rocks all result in methane release on Earth (e.g., \cite{formisano2004detection}). 
Also, methane clathrates release methane into the atmosphere once unlocked from the melting of ice sheets, lenses, and frozen ground \citep{prieto2006interglacial}.
A key question to be addressed is which one of these phenomena (or perhaps all) contributed to the production of the detected methane on Mars? 

On Earth, arrays of methane detectors have been in usage for decades. Recent advances include quantum optimal detectors for monitoring industrial gas leaks. Earth-based methane investigations include long-range, high-sensitivity, and high-speed quantitative imaging of the shape and concentration of particulates in gas plumes.  

For planetary landers on Mars, methane detectors have been developed and deployed for both orbiting and roving spacecraft.
Low concentrations (up to 10s ppb) of methane have been controversially detected in the modern Martian atmosphere through Earth-based telescopes \citep{mumma2004detection,krasnopolsky2004detection}, and by the Mars Express spacecraft \citep{formisano2004detection}. Viking Landers 1 and 2 detected chloromethane at 15 ppb and .04-40 ppb, respectively [e.g., \cite{klein1992search}] by using chemical-reaction experiments.
The Curiosity rover in Gale crater has recently identified variable methane concentrations up to 21 ppb, using the Sample Analysis at Mars (SAM) instrument \citep{webster2015mars}, albeit in what appears to be transient releases.
Smart quantum methane sensor systems provide the next step in the exploration and detection of methane sources.

Individual sensors only provide a point source detection, which makes it extremely challenging to pinpoint the spatial source of methane gas emissions; in contrast, deploying a sensor array would provide uniquely valuable data to understand the methane cycle on Mars.

\subsubsection{Distributed Instrument Configuration and Strength}
\label{sec:sciencecase:tracegas:cs}

Regional and global-scale methane plumes can be detected with orbiters and telescopic observations; once a region where emissions potentially originate from is identified, a small number of detectors, as part of a distributed network, can be deployed to pinpoint the origin of emission. If an orbiter is used to detect large plumes, a particularly attractive concept of operation would entail the orbiter performing remote observations and carrying a number of surface sensing units; once a plume is detected in a given region, the orbiter could deploy the sensing units to the surface in the region of interest to obtain an accurate localization.
Regions of interest include fissures and possible serpentinization sites. For instance, four methane detectors comparable to Curiosity's SAM Tunable Laser Spectrometer (TLS) \citep{mahaffy2012sample}, positioned around the edge of Gale crater over a distance of tens to hundreds of kilometers, could potentially lead to location of a plume source originating from within the crater, such as the one observed by Curiosity in 2015 \citep{webster2015mars}. %

\begin{figure}[h]
\centering
\includegraphics[width=.9\textwidth]{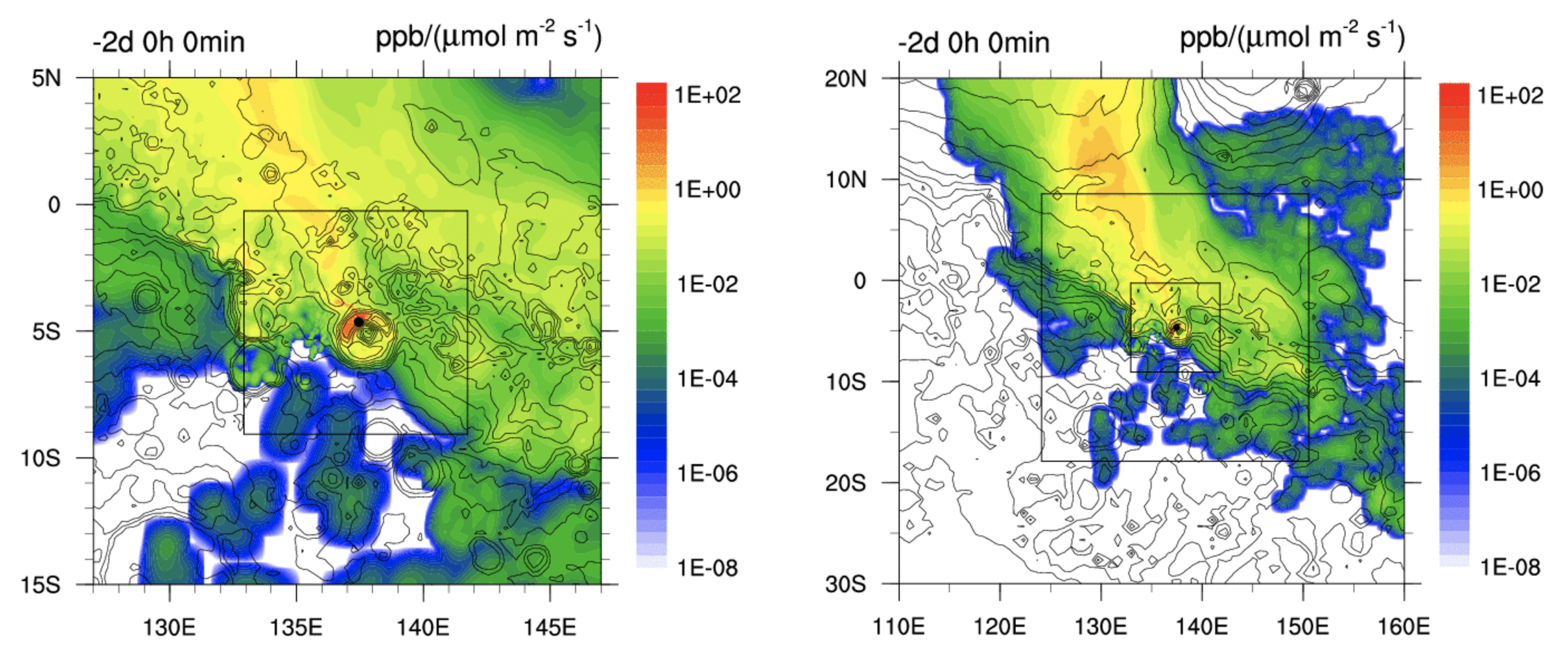}
\caption{Identification of the source of methane emission sites relies on accurate knowledge modeling of atmospheric circulation. In the figure, the MarsWRF general circulation model \citep{MarsWRF} is used to resolve the local atmospheric circulation at Gale crater with 4 km resolution, and the Stochastic Time-Inverted Lagrangian Transport (STILT) model, adapted to Mars, is used to trace, backwards in time, upstream locations from where surface emissions can exert influence on the detected methane concentration. From \cite{mischna2020gasdetection}.}
\label{fig:science:methane:detection}
\end{figure}

Localizing the source of a gas plume requires accurate knowledge of atmospheric circulation [\cite{mischna2020gasdetection} and Figure \ref{fig:science:methane:detection}]. High-quality mesoscale models of planetary atmospheric circulation are available \citep{MarsWRF}; however, as discussed in Section \ref{sec:sciencecase:climate}, knowledge of ground-truth information is critical to the accuracy of these models. To this end, augmenting a trace gas detection distributed instrument with a large number of low-cost, high-TRL weather stations modeled after those flown on Curiosity, InSight, and Perseverance \citep{gomez2012rems,banfield2019insight,rodriguez2021mars} distributed in the region of interest could greatly increase the accuracy of the distributed instrument, in addition to achieving many of the goals discussed in Section \ref{sec:sciencecase:climate}. %

\subsubsection{Unique Technical Challenges}

\paragraph{Advance knowledge and potential precursor missions.}

\revmm{Observations from the ExoMars Trace Gas Orbiter have not revealed the presence of global-scale gas plumes, and recent publications \citep{Fonti15TES} have called into question the accuracy of past detections, suggesting that Martian methane emissions are likely to be local in scale.}  Recent work \citep{Webster2021methane}, suggests confinement of the methane signal largely to within Gale crater from diurnal growth and collapse of the planetary boundary layer.
A distributed instrument for gas detection is inherently a regional (as opposed to global) instrument, and its added value lies in the detection and localization of the source of local plumes; therefore, advance knowledge of areas where gas emissions are likely is critical to target instrument deployment.
Curiosity's observations of gas plumes and movements of air are local to Gale crater; precursor \revI{detection of trace gases} will be required to motivate deployment at other regions of interest.

\paragraph{Reliance on atmospheric circulation models.} Localization of the source of plumes requires precise estimation of atmospheric transport, which is influenced by local wind. To reduce the uncertainty of transport models, adding a wind gauge to individual landed sensors would be highly beneficial; an even more beneficial addition would be to include a large number of lightweight, low-cost weather stations to be deployed in the area of interest, as discussed above.

\subsection{Common challenges}
\label{sec:sciencecase:common-challenges}

\revmm{We conclude this section by discussing two challenges that broadly affect all outlined concepts: cost-effective delivery of the sensors to the planetary surface, and long-term survivability.

\subsubsection{Cost-Effective Delivery} For many distributed instruments, basic sensor costs do not present a significant hurdle, nor do they offer technology development challenges; rather, there remain unresolved challenges concerning methods of deployment—how the sensor payloads are to be carried and deployed across the planet — and the carrier system by which the sensors are delivered. 
Despite the small nature of each individual sensor, costs associated with a ‘‘mothership’’ or payload carrier may increase mission costs for a dedicated mission. These concerns are especially relevant for instruments that require global distribution, e.g., those targeted to climate and magnetometry investigations. The requirement to deploy across a range of latitudes and longitudes presents non-trivial questions about how the sensors are to be released—whether during approach to the body of interest, or by an orbiting carrier.  This uncertainty requires additional analysis, and the best solution may present additional costs to a notional stand-alone mission design.  An alternative approach, deployment as a series of secondary, ‘‘ride-along’’ SmallSat payloads, also requires further study.

\subsubsection{Long-Term Survivability} Individual elements, with low size, weight, and power (SWaP), must also manage potentially challenging environmental conditions; it is also necessary to ensure a sufficiently long lifetime for each sensor package.  
For all the science concepts of interest, multi-year investigations are highly desirable to maximize the likelihood of detecting events of interest and to capture seasonal or interannual variability.
For instance, a strawman threshold mission investigating the Martian climate would last a minimum of one Mars year, so as to fully cover the three primary Martian climate cycles.  Ideally, though, a mission lifetime of several Mars years is more desirable so as to capture as much of the interannual variability in the climate cycles as possible: while CO$_2$ and water cycles are mostly consistent year over year, the dust cycle is highly variable and unpredictable; e.g., \cite{zurek1993interannual}.
Likewise, a mission to investigate seismic activity should perform monitoring over a long time horizon; for reference, InSight's reference mission was baselined to last one Martian year (corresponding to approximately two Earth years), and was subsequently extended by an additional Martian year.
Similarly, a distributed instrument designed to observe and localize trace gas plumes should last months to years, so as to increase the likelihood of observing and, critically, localizing a meaningful number of plumes.
Survivability concerns are especially critical for instruments requiring global coverage, e.g., climate investigations.
High latitude (polar) elements of the distributed instrument may experience extended periods of low temperature and darkness throughout the polar winter.  This can, in part, be mitigated by deployment of sensors only to mid- or low latitudes, but to enable true global coverage, it is necessary to deploy elements to polar latitudes, and therefore to keep elements of these higher latitude payloads and their electronics warm and energized.  This will impact payload design and SWaP of the payload.

An attractive concept of operations to overcome these limitations is for the sensors to take measurements at slow cadence and analyze the data through simple on-board detectors (e.g., pressure wave detectors for climate investigations, plume detectors for trace gas instruments, and seismic activity detectors for seismology); when a sensor detects an event of interest (e.g., a methane spike), all sensors in the distributed instrument switch to a high-rate mode to characterize its spatial and temporal evolution. Such a concept of operations requires  development of both on-board autonomy technology to detect event of interest, and sensor-to-sensor communication protocols where an orbiter autonomously relays data between nodes.

}
 
\section{Enabling Technologies}
\label{ch:technologies}

Realizing the benefits of distributed instruments poses novel technology challenges that are not typically faced by monolithic instruments and missions. In this section, we survey the state-of-the-art in ancillary technologies uniquely relevant to distributed instruments, namely:
\begin{itemize}
\item Localization and synchronization, i.e., how to determine the spatial and temporal correlation of sensor measurements;
\item Communications, i.e., how to relay data from the sensors to scientists, and how sensors can trigger each other;
\item Adaptive sampling and science autonomy, i.e., how to intelligently trigger the sensors and compress data so as to fit within tight power and bandwidth constraints; and
\item Coordinated sensor placement, i.e., how to \revmm{efficiently achieve the desired spatial distribution of the sensors} on the planetary surface.
\end{itemize}

In addition, individual sensors of a distributed instrument are likely (although not necessarily required) to fit in a small SWaP envelope, to allow for efficient deployment of the instrument. Accordingly, the design of distributed instruments is likely to be constrained by technical challenges that affect small instruments, namely:
\begin{itemize}
\item On-board computation and storage, i.e., what hardware is available to process and store data on board;
\item Power, i.e., how to ensure the sensors can operate for the entirety of the required mission; and
\item Thermal control, i.e., how to satisfy the thermal constraints of the sensors and ancillary technologies.
\end{itemize}

At the end of this section, we also compare the technology readiness level of state-of-the-art technologies with the scientific requirements discussed in Section \ref{ch:sciencecase}, and identify areas where additional technology development is required to bring distributed instruments to a level of technology readiness compatible with spaceflight.

\subsection{Localization and Synchronization}
\label{sec:technologies:pnt}

Spatial localization and time synchronization of individual sensors is required to provide scientific context for the sensor readings; it also enables synchronization of readings, which is critical for applications such as seismology, and can enable significant power savings by allowing individual nodes to only attempt to communicate with an orbiter when it is overhead.

The problems of localization and synchronization are closely intertwined, and also closely tied to the selected communications architecture.

\subsubsection{Time Synchronization}
\label{sec:technologies:pnt:sync}

A number of high-TRL, space-qualified solutions for on-board timekeeping are available, including chip-scale atomic clocks that provide extremely low drift over long periods of time; if the clocks of individual sensors must be kept synchronized to nanosecond-level accuracy, RF-based synchronization techniques can be used to synchronize the individual sensors' oscillators, countering relative drift.
Table \ref{tab:pnt:timing} reports key performance parameters, SWaP, and TRL for a variety of representative timing and time synchronization solutions applicable to distributed instruments.

\paragraph{On-board clocks}

A large variety of high-TRL, space-rated oscillators are available \citep{vectron_space_oscillators, vectron_1404_oscillator} that offer low power consumption and comparatively low frequency drift in the range of 5 parts per million per year (equivalent to 2.5 minutes/year). Temperature-controlled crystal oscillators (TXCO) \citep{Vectron_2105_oscillator} and oven-controlled crystal oscillators (OXCO) \citep{vectron_OX249_oscillator} offer higher precision (with five to ten times less drift for a given temperature) and, crucially, much lower sensitivity to temperature, at the price of increased size, weight, and power.
For very low drift applications, chip-scale atomic clocks \citep{microsemi_csac} can provide an additional order-of-magnitude improvement in drift performance in a package with size comparable to OXCOs;  space-qualified models are available.

\paragraph{RF-based synchronization}
Even the most accurate local oscillators exhibit some drift with respect to one another over sufficiently long timescales. %
RF-based synchronization techniques can be used to re-synchronize multiple oscillators to a common source, e.g., an orbiter's on-board oscillator. Commercial, military-grade synchronization solutions are available  \citep{ensco_pnt} that provide real-time sub-ns synchronization over radio links; in the space domain, an ad hoc Time Transfer System (TTS) has been demonstrated as part of the GRAIL mission \citep{klipstein2013lunar}. %

\begin{deluxetable}{rcccccccc}
\caption{Comparison of time synchronization solutions}
\label{tab:pnt:timing}
\tablehead{
Solution family & \rotatebox{90}{\shortstack{Short-term accuracy /\\1 s Allan deviation}} & \rotatebox{90}{Long-term drift} & \rotatebox{90}{Relative synchronization} & \rotatebox{90}{Size} & \rotatebox{90}{Weight} & \rotatebox{90}{Power} & \rotatebox{90}{Space-qualified} & \rotatebox{90}{TRL}}
\startdata
XO & $\sim 10^{-10}$ & 5 ppm/yr & No & 13x10x3 mm & 3 g & 0.03 W & Yes & 9 \\
TXCO  & $\sim 10^{-10}$  & 1 ppm/yr & No & 20x20x3 mm & 10 g & 0.05 W & Yes & 9 \\
OXCO  & $8\cdot 10^{-12}$ & 0.4 ppm/yr & No & 35x20x10 mm & 14 g & 2 W & Yes & 9 \\
CSAC  & $3\cdot 10^{-10}$ & 0.1 ppm/yr & No & 40x35x11 mm & 35 g & 0.12 W & Yes & 7 \\
COTS RF synchronization  & 0.1 ns & 0 & Yes & 175x77x34 mm & 400 g & 5 W & No & - \\
Ad hoc RF synchronization & 100 ns & 0 & Yes & - & - & - & Yes & 9 \\
\enddata

\tablerefs{\cite{vectron_1404_oscillator,Vectron_2105_oscillator,vectron_OX249_oscillator,microsemi_csac,ensco_pnt,klipstein2013lunar}}
\end{deluxetable}

\subsubsection{Localization}

If the sensor is statically positioned by a larger vehicle whose location is well-known (e.g., it is placed by a rover's robotic arm) its location can be accurately computed based on the location of the carrier vehicle.
Conversely, if the sensors perform EDL independently or are delivered dynamically (e.g., they are ejected from a larger vehicle), two-way RF ranging can provide extremely accurate localization with respect to other sensors and to orbiters; and one-way Doppler-based ranging can provide less accurate estimations from a single orbiter without the need for a radio transponder. Visual odometry-based techniques and terrain-relative navigation are routinely used for localization of large assets, but they generally require comparatively high-SWaP sensors, which makes them less attractive for distributed instruments.

Table \ref{tab:pnt:localization} reports key performance parameters, SWaP, and TRL for representative localization solutions, discussed in more detail next.%

\paragraph{RF-based ranging and triangulation}

RF-based ranging can provide relative ranging with extremely high accuracy relying on either a well-characterized radio transponder, or a well-synchronized timing source.
In \emph{two-way} ranging, the transmitter sends a signal, and the receiver's transponder retransmits the signal back; by computing the time between transmission and reception, and given the known performance characteristics of the transponder, the transmitter computes the relative range with accuracy limited only by the temporal resolution of the measuring equipment.
In \emph{one-way} ranging, the transmitter broadcasts a signal at an agreed-upon time. The transmitter and the receiver's clocks are assumed to be synchronized (as discussed in Section \ref{sec:technologies:pnt:sync}). The receiver compares the reception time with the agreed-upon transmission time, and computes the range from the transmitter based on the time-of-flight.

In both one-way and two-way ranging, by computing multiple ranges between each sensor and an orbiter, the location of a sensor can be triangulated to an accuracy comparable to the knowledge of the orbiter's location in orbit. %

A variety of solutions for RF-based localization have been demonstrated for Earth applications, ranging from inexpensive COTS solutions for localization in small volumes \citep{Jimenez2017Decawave} to military solutions that provide sub-cm accuracy over ranges of tens of kilometers \citep{ensco_pnt}. 
However, to the best of the authors' knowledge, existing solutions are \emph{not} rated for space environments; existing communication equipment for space proximity links generally does not provide proximity ranging functions \citep{CheungLeeEtAl2019}, although ad hoc solutions have been developed for the GRACE \citep{davis1999grace} and GRAIL \citep{klipstein2013lunar, weaver2010performance} gravimetry missions.
 
\paragraph{Doppler-based localization from an orbiter}

Doppler-based localization observes the Doppler shift of received signals and uses the known ephemerides of either the transmitter or the receiver to estimate the relative location of the two. Doppler-based ranging has been widely used for localization on Earth, including for search-and-rescue applications \citep{argos_system} where a satellite can compute the location of a surface transmitter with accuracy of tens of meters during a single overpass  \citep{ellis2020use}. 
Surface receivers equipped with an accurate on-board oscillator can also compute their own location relative to an uncooperative orbiter \citep{CheungLeeEtAl2019}; availability of surface reference stations with known location (e.g., a lander) can increase accuracy to a few meters.

\paragraph{Image-based localization at deployment}

Terrain-relative navigation (TRN) has been proposed for landing on rocky bodies including Mars \citep{johnson2012lander,johnson2017lander}, the Moon \citep{JohnsonMontgomeryAero08}, and Titan \citep{MattiesDafrtyEtAlAero2020}, and was used for precision landing on the Perseverance rover \citep{johnson2017lander,JPL2019TRN}. TRN can provide accurate, real-time localization with respect to a known terrain map during descent and landing; however, it requires dedicated downward-facing cameras and powerful on-board processing. The key benefit of TRN is the ability to provide real-time localization during the dynamic descent and landing phase, enabling precision landing - a capability that may not be needed by all distributed instruments.

\paragraph{Attitude determination}

If it is required to characterize the attitude and orientation of the sensor package, a number of techniques are available.  Sun sensors can be used to compute the full attitude of the sensor unit to a precision of a few degrees, if a time reference is available \citep{LambertFurgaleEtAl2012}.
 For bodies with a magnetic field, a magnetic sensor can be used to constrain the attitude of the sensor.
 An IMU can provide information about roll and pitch, although a Sun sensor or magnetic sensor is typically required to determine yaw.
Finally, if a camera sensor is available, image-based localization can be used to constrain the sensor's attitude with respect to local visible landmarks.

\begin{deluxetable}{rcccccccc}
\caption{Comparison of localization solutions}
\label{tab:pnt:localization}
\tablehead{
\colhead{Solution family} & \colhead{\rotatebox{90}{Relative accuracy}} & \colhead{\rotatebox{90}{Absolute accuracy}} & \colhead{\rotatebox{90}{Time to localize}} & \colhead{\rotatebox{90}{Size}} & \colhead{\rotatebox{90}{Weight}}& \colhead{\rotatebox{90}{Power}} & \colhead{\rotatebox{90}{Space-qualified}} & \colhead{\rotatebox{90}{TRL}}
}
\startdata
COTS RF-based ranging & 5 m & - & 3 ms & 175x77x34 mm & 400 g & 5 W & No & -\\
Ad hoc RF-based ranging  & $\sim$ 1 nm & - & - & - & - & - & Yes & 9\\
Doppler-based localization  & 50 m & 50 m & 3 s & - & - & - & Yes & 9\\
Image-based localization & - & 40 m & 10 s & 6U chassis & $\sim$ 10 kg & $\sim$ 100 W & Yes & 8\\
\enddata
\tablerefs{\cite{ensco_pnt,klipstein2013lunar,CheungLeeEtAl2019,argos_system,johnson2012lander,johnson2017lander}}
\end{deluxetable}

\subsection{Communications}
\label{sec:technologies:communications}

The communication subsystem is a critical part of a distributed instrument. Not only is it required to collect data from individual sensor units, but it can also allow sensors to communicate with each other in response to sensed events, enabling reactive and agile sensing and potentially reducing power requirements. In addition, the communication subsystem is essential to localization and time synchronization, as discussed in Section \ref{sec:technologies:pnt}.

In this section, we survey existing communication equipment suitable for distributed instruments, ranging from NASA-developed transponders and radios, to commercial radios for SmallSats, to COTS radios for Earth applications.

\subsubsection{A typical communications architecture from planetary surfaces to Earth}

\begin{figure}[h]
\centering
\includegraphics[width=.6\textwidth]{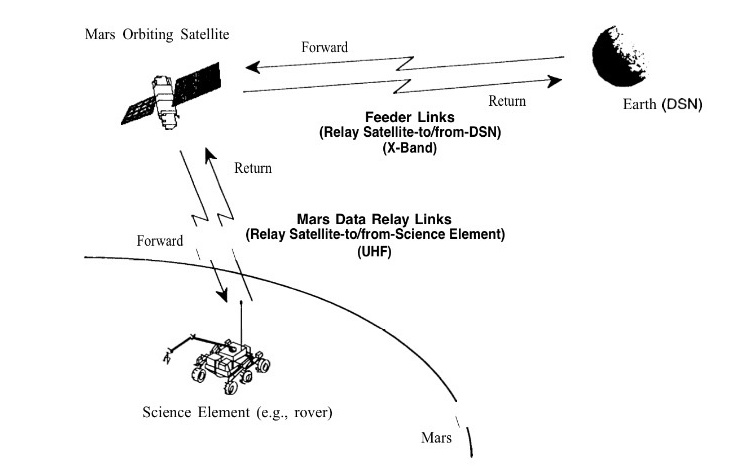}
\caption{A typical relay communication architecture. Ground-based assets communicate with an orbiter over UHF links. The orbiter, in turn, communicates to Earth over X-Band. From \cite{horne1997telecommunications}.}

\label{fig:technologies:communications:architecture}
\end{figure}

Figure \ref{fig:technologies:communications:architecture} shows a typical communication architecture for communications from a surface asset to Earth, mediated by an orbiter. The surface asset is equipped with a UHF L-band radio and an omnidirectional antenna; an orbiter communicates with the surface asset through a directional antenna, and relays data to and from Earth with a high-gain X-band or K-band transceiver. L-band is typically used for surface-to-orbit communications because it offers favorable propagation characteristics through planetary atmospheres. 
The typical distance between the surface asset and the orbiter is in the order of hundreds of kilometers; however, standoffs of up to sixteen thousand kilometers have been demonstrated by Galileo's entry probe on Jupiter, and the Cassini-Huygens probe on Titan.

\subsubsection{NASA-developed Transponders}

The NASA Jet Propulsion Laboratory has developed a number of high-performance, low-SWaP transponders for deep space applications,%
including the Iris X-band transponder flown on the MarCO CubeSats to Mars, the Universal Space Transponder (UST) family \citep{Pugh2017}, and the UST-Lite transponder \citep{kobayashi2023ust}, which offers multi-band capabilities, high uplink and downlink rates, and very high sensitivity in a 3 kg, 2700 cm$^3$ package.

Table \ref{tab:technologies:communications:jpl} reports a comparison of key performance parameters of NASA-developed deep-space transponders. We refer the reader to \cite{Kobayashi2018} for a detailed review of NASA-developed transponder technologies for SmallSats.

\begin{deluxetable}{rcccc}
\caption{Comparison of JPL Deep-Space Transponders. From \cite{Kobayashi2018}.}
\label{tab:technologies:communications:jpl}
\tablehead{
\textbf{Radio Specification} & \textbf{Units} & \textbf{Iris} & \textbf{UST-DS} & \textbf{UST-Lite}
}
\startdata
Frequency Bands& - & X up, X down & S/X up, S/X down & X/Ka up, X/Ka down \\
&&&simultaneous dual band & simultaneous dual band \\
Mass& kg& 1.0 & 5.4 & 3.0 \\
Volume& cm$^3$ &600&7500&2700\\
Bus Input Voltage&Vdc&9 – 28&22 – 36&22 – 36\\
DC Power&W&16&45&30\\
Processors&&Xilinx V6 + LEON-3FT&Xilinx V5 + SPARC V8&Xilinx V5 + LEON-3FT\\
Receiver Noise Figure&dB&3.5&2.1&2.1\\
Receiver Sensitivity&dBm&-151 @ 20 Hz LBW & -160 @ 20 Hz LBW  & -160 @ 20 Hz LBW\\
Uplink Rate & sps & 62.5– 8k & 7.8125 – 37.5M & 7.8125 – 37.5M\\
Downlink Rate & sps & 62.5– 6.125M & 10 – 300M & 10 – 300M\\
Telemetry Encoding &  & Conv, RS, Turbo & Conv, RS, Turbo, LDPC & Conv, RS, Turbo, LDPC\\
Radiation Tolerance (TID) & krads & 23 & 50 & 300\\
S/C Interface &  & 1 MHz SPI & 1553, SpaceWire, RS-422 & 1553, SpaceWire, RS-422 \\
\enddata

\end{deluxetable}

The SWaP of these radios is optimized for larger surface and orbital assets, and is amenable to further reductions; however, this has not proved necessary for current and planned NASA missions, where the radio system SWaP has not been a major bottleneck.

\subsubsection{Commercial SmallSat Radios}

A number of low-power radios for SmallSats are available from commercial vendors. As a representative example, Vulcan Wireless \citep{Vulcan2020} and Rincon Research \citep{Rincon2020} both provide software-defined radios with a mass of hundreds of grams and a volume of hundreds of cubic centimeters, %
 designed to fit in the demanding SWaP envelope of CubeSats, with flight heritage in the low-Earth orbit environment. Vulcan Wireless's radios have also been selected by NASA for development in support of the Artemis lunar exploration program \citep{VulcanWirelessSAA}.

\subsubsection{COTS radios}

Commercial off-the-shelf (COTS) radios for automotive-grade applications are attractive for highly miniaturized purposes, offering low power consumption in highly compact packages; however, COTS technologies in this class are generally optimized for  low-range, high-bandwidth applications, requiring significant adaptations to perform well over the long standoff distances required in spaceflight.

\begin{figure}[h]
\centering
\ifaaspsj
\includegraphics[width=.42\textwidth]{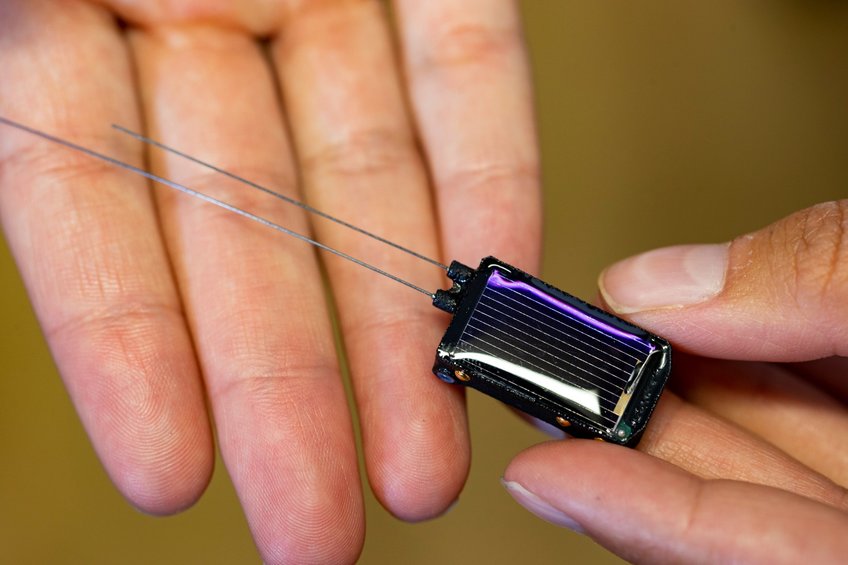}
\else
\begin{subfigure}[b]{.28\textwidth}
\includegraphics[width=\textwidth]{communications/CC3200.jpg}
\caption{TI CC3200}
\end{subfigure}
\begin{subfigure}[b]{.35\textwidth}
\includegraphics[width=\textwidth]{communications/ICARUS.jpg}
\caption{ICARUS transmitter architecture. Image credit MPI f. Ornithology/ MaxCine }
\end{subfigure}
\begin{subfigure}[b]{.28\textwidth}
\includegraphics[width=\textwidth]{communications/icarus_in_hand.jpg}
\caption{ICARUS transmitter. Image credit MPI f. Ornithology/ MaxCine}
\end{subfigure}
\fi
\caption{ICARUS COTS ultra-low-SWaP transponder. Image credit MPI f. Ornithology/ MaxCine}
\label{fig:technologies:communications:cots}
\end{figure}

A standout solution in this space is represented by the ICARUS system developed by Max Planck Institute for animal tracking \citep{ICARUS}. ICARUS (shown in Figure \ref{fig:technologies:communications:cots}) consists of a 5-gram unit equipped with a radio, a GPS payload, and a power system. Thousands of ICARUS units communicate with an orbiter, which can also address commands to individual radio units. To date, around five thousand ICARUS units have been demonstrated in the field, supported by an orbiting receiver hosted on the International Space Station \citep{ICARUS_projects}. 
Deployment of a SWaP-optimized, 1U receiver as part of a larger 16U CubeSat is planned for 2024 \citep{ICARUS_cubesat}.
Each radio unit on the ground knows its location and the ephemerides of the orbiter, and only turns on when the orbiter is overhead; thanks to this aggressive power management, ICARUS sensors are able to last for multiple years in the field with no maintenance. Table \ref{tab:technologies:communication:icarus} shows preliminary performance parameters for the ICARUS system.

\begin{table}[h]
\centering
\caption{Specifications of the ICARUS radio system \citep{ICARUS}.}
\label{tab:technologies:communication:icarus}
\begin{tabular}{r|l}
Frequency Band & VHF \\
Encoding & CDMA \\
Range & 800 km\\
Uplink bandwidth& 1.5 MHz \\
Downlink bandwidth & 50 kHz \\
Data rate & << 500 bps \\
TX duration & 35 s \\
TX power (ground) & 6 mW \\
TX power (space) & 100 W (unoptimized) \\
Payload size & 200 bytes per pass \\
Weight & 5 g
\end{tabular}

\end{table}

\subsubsection{Communication Protocols}

In typical spaceflight applications, the communication topology is straightforward: surface assets communicate to an orbiter which, in turn, relays the information to Earth, and vice versa.
In contrast, in distributed instruments, individual sensors may wish to communicate with each other to perform autonomy functions (discussed in Section \ref{sec.AdaptiveSampling}), e.g., to trigger follow-on observations in response to an event of interest observed by a single sensing unit.
The flight-proven delay- and disruption-tolerant networking (DTN) communication protocol \citep{burleigh2003dtn} can be used to perform multi-point communications in an environment where individual sensors are unable to directly talk to each other, and an orbiter is only available sporadically.

\subsection{Adaptive Sampling and Science Autonomy}
\label{sec.AdaptiveSampling}

\subsubsection{Science at a distance}
Commanding a network of sensors to  take coordinated, spatially separate recordings as one instrument will bring new operations challenges. We will explore two core challenges here.

First, returning the complete record of science data will likely be challenging or infeasible. Missions using distributed instruments are likely to collect significantly more data compared to monolithic missions, especially for applications such as seismometry. Additionally, the impending increase in crewed exploration to the Moon and Mars will further ration DSN availability \citep{Deutsch2016}. If a distributed instrument generates data faster than it can transmit -- a likely scenario given these considerations -- there must exist a way to intelligently distill that information to remain within bandwith limits. 

Second, radio signals take on the order of minutes to hours to cross the interplanetary distances between Earth and planetary surfaces of interest, which imposes a serious hurdle if ground teams need to  react quickly to time-varying phenomena. An Earth-in-the-loop communication cycle requires time to downlink information, analyze the data from the many sensor nodes, and uplink commands in response. Thus, if there is an event short enough to encroach upon this time window, the distributed instrument must react on its own, or risk missing scientifically valuable data.

\subsubsection{Opportunities for science autonomy}
On-board science autonomy -- software capable of intelligently sampling, selecting, or prioritizing scientific data -- is one tool capable of mitigating both bandwidth limitations and communications delays  \citep{azari2020integrating, Theiling2022_science_autonomy} through \emph{knowledge compression}, \emph{data prioritization} and \emph{reactive observation} of events of interest. %

\paragraph{Knowledge compression}

Knowledge compression goes beyond classic data compression algorithms (e.g., JPEG or ZIP) by excising and summarizing the most \emph{scientifically valuable} representation of the raw data. On-board science instrument autonomy software for JPL’s Ocean Worlds Life Surveyor (OWLS; \cite{wronkiewicz2023onboard, mauceri_2022}) is one example of a project relying on knowledge compression. Several microscopes aboard the OWLS project record video of water samples. On-board software identifies life-like particles that appear to swim under their own power, and summarizes their appearance and movement. By transmitting a spatiotemporal time series (i.e., \textit{x}, \textit{y}, and time points) and a single image crop of life-like particles, the software can compress scientifically valuable information by over three orders of magnitude compared to the raw videos. Another example is the Content-based On-board Summarization to Monitor Infrequent Change (COSMIC; \cite{Doran2020}) project. COSMIC is aimed at detecting changes in a remote sensing data stream that is being continually acquired, e.g., monitoring for large surface changes like fresh impacts or seasonal polar cap variations aboard a Martian orbiter (e.g., the Mars Reconnaissance Orbiter). Rather than downlink all collected data, COSMIC prioritizes the transmission of orbital images containing the most scientifically interesting surface changes.

\begin{figure}
	\centering
	\includegraphics[width=0.6\textwidth]{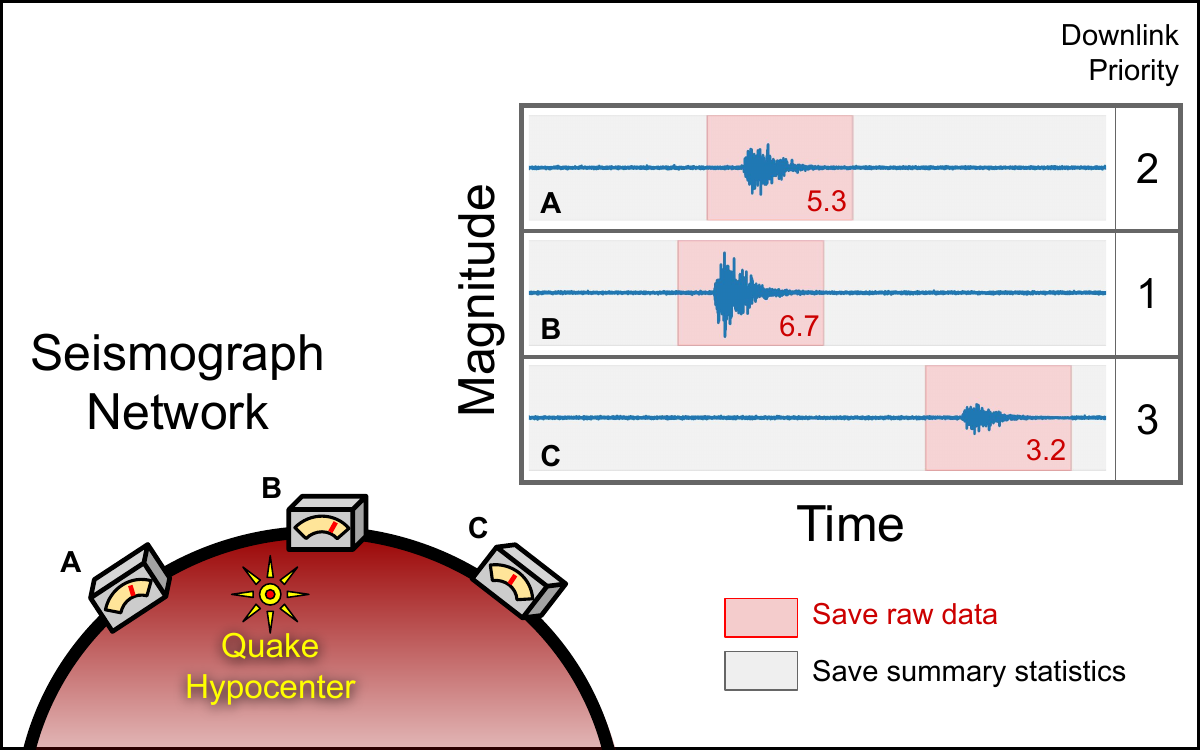}
	\caption{Autonomy will facilitate selection and downlink prioritization of scientifically valuable data. Left: Three spatially separated seismographs record ground motion from quakes, impacts, or other events that produce ground motion. Right: The raw seismograph data will contain rare, scientifically interesting events (e.g., quakes) against a background of less interesting data. Autonomy allows each instrument to detect seismic onsets and then select a window of raw data (red) for return to ground. Data from other times (light gray) can be summarized with simple statistics (e.g., mean and variance). Furthermore, the quake magnitude (red numbers) can be used to determine downlink priority (far right).}
	\label{fig.autonomy_compression_prioritization}
\end{figure}

On distributed instruments, knowledge compression will represent a valuable capability as operating a network distributed sensors can create substantially more raw data than previous monolithic missions. For example, a distributed network of Mars seismographs might offer a more holistic view of Mars’s internal dynamics and impact events (shown in Figure \ref{fig.autonomy_compression_prioritization}). However, large seismic and impact events are rare \citep{CEYLAN2022}. Science autonomy could identify scientifically interesting time intervals (i.e., when a seismic event occurred) and limit data transmission to the raw recordings around the time of the events. Data outside that range can be summarized and transmitted as simple statistics (e.g., mean and standard deviation) to save valuable data bandwidth.

\paragraph{Data Prioritization}
Rank ordering data is a separate technique useful for triaging the most valuable data to downlink. For scientific targets that are especially far away (e.g., Jupiter, Saturn, or Uranus) or residing in environments known to quickly destroy hardware (e.g., Venus and the moons of Jupiter), properly ordering data for transmission can reduce the risk that crucial science data is collected but not transmitted to ground teams. Therefore, generating estimates of scientific utility for each parcel of data is an important capability for an autonomous distributed instrument. Continuing the previous example, an autonomous distributed instrument could downlink data from seismic events in an order determined by their magnitude --- a simple proxy for scientific utility. Successfully implementing this type of capability requires that scientists and software engineers collaborate to develop a prioritization algorithm that makes reasonable and transparent decisions. A generalized software platform is underway at NASA to permit simultaneous data prioritization for multiple sensors \citep{Doran_SYNOPSIS}.

\paragraph{Reactive systems}

Some phenomena transpire on timescales incompatible with ground commanding, e.g., transient weather events such as dust devils or trace gas plumes. On-board science autonomy could permit near-instantaneous reactions, making it well-suited to detect and respond to ephemeral events.

The Opportunity rover deployed one such autonomous algorithm to identify dust devils \citep{Castano2008}. The system analyzed temporal sequences of images and identified large blobs of pixel changes, which corresponded primarily to dust devil events. Opportunity then prioritized these images for return. The Curiosity and Perseverance rovers also use an autonomous algorithm called AEGIS (Autonomous Exploration for Gathering Increased Science) to identify nearby geologic targets for ChemCam sampling \citep{Franciseaan4582}. AEGIS allows the rover to autonomously select ChemCam targets immediately after a drive, before the ground team has analyzed the rover’s surroundings. Importantly, the AEGIS autonomous system is a collaboration between scientists and autonomy experts, with the final tuning carried out by mission scientists.

\begin{figure}[h]
    \centering
	\includegraphics[width=0.45\textwidth]{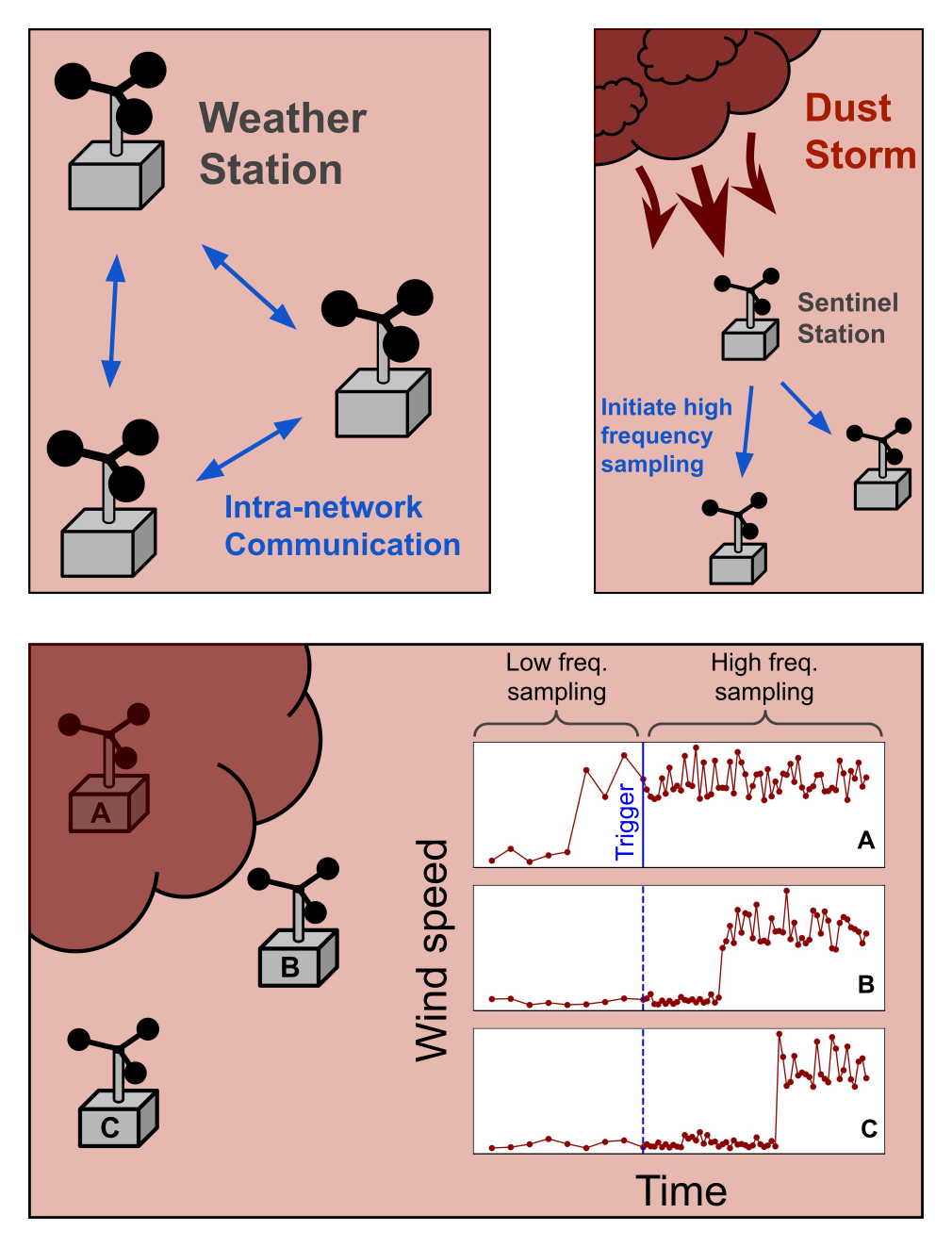}
	\caption{Autonomy will allow distributed instruments to increase science return by detecting and reacting to rare events while preserving system resources. Top left: Three spatially separated weather stations regularly communicate with each other. Top right: A dust storm approaches and is detected by one station. The event triggers a network response to initiate high frequency data collection across all stations. Bottom: Example data collection from all three weather stations before and during a dust storm. By default, the network records wind speed at low rates to conserve energy and data volume. However, once a rare weather event is detected at Station A, the distributed instrument increases its sampling frequency at all locations to capture the approaching storm in detail.}
	\label{fig.autonomy_reactive_systems}
\end{figure}

Distributed instruments are particularly well-suited to react to transient phenomena because any one individual sensor can act as a ``sentinel'' and trigger a reaction from the broader network. With gas plumes, for example, if a sensing unit detects anomalous changes in the atmospheric composition, it can instruct all nodes of the distributed instrument to increase their sampling rate to capture the gas diffusion pattern in detail. Martian dust storms are another potential scientific target of distributed instruments. Many of these events are small and short lived while a few grow into much larger storms, and this transition is poorly understood. If each component of the distributed instrument could detect deviations in wind speed, atmospheric pressure, or suspended dust, it could instruct other, spatially separate sensors in the distributed instrument to start spending more on-board resources on data capture in anticipation of the approaching storm front (Figure \ref{fig.autonomy_reactive_systems}). The resulting high resolution in-situ data would be valuable for informing future atmospheric studies and models. 

\subsubsection{Future directions}

Science autonomy holds promise to empower scientific discovery by mitigating key limitations related to data bandwidth, adaptive sampling, and response times. However, it remains a relatively new technology that requires more development for some of the aforementioned scenarios.
To achieve this vision, three key directions for future research are of interest.
First, a tight collaboration is needed between the scientists who will consume the data from distributed instruments and the software team crafting the autonomous system to capture it. The best outcomes will occur when scientists are involved in the development of this technology as well as its operation during the mission. Creating the culture and tools to achieve this environment will require more research. 
Second, it is critical for autonomous systems to be explainable and transparent, in order to be worthy of the scientists’ trust \citep{Mandrake2022}. Such systems must convey information on how the system arrives at consequential decisions (e.g., why one data set was returned before another), and provides context around important data points (e.g., several minutes of background recordings before and after a large quake). Toward this goal, algorithmic complexity (e.g., black box vs. glass box models) must be evaluated to identify the appropriate tradeoff between performance and transparency.
Finally, the autonomous system should fail gracefully. It’s infeasible to predict every situation an autonomous distributed instrument might encounter. Therefore, the system must have built in safety valves to permit manual overrides after undesirable autonomous decisions. The system should also include validation checks to recognize when it is operating outside normal conditions and then refrain from making decisions that could jeopardize the mission’s science return (e.g., detect if one sensor in an instrument suite is broken, to prevent the squandering of precious bandwidth on its tainted data).

\subsection{Sensor Placement}

\begin{figure}[!h]
\centering
\includegraphics[width=0.5\textwidth]{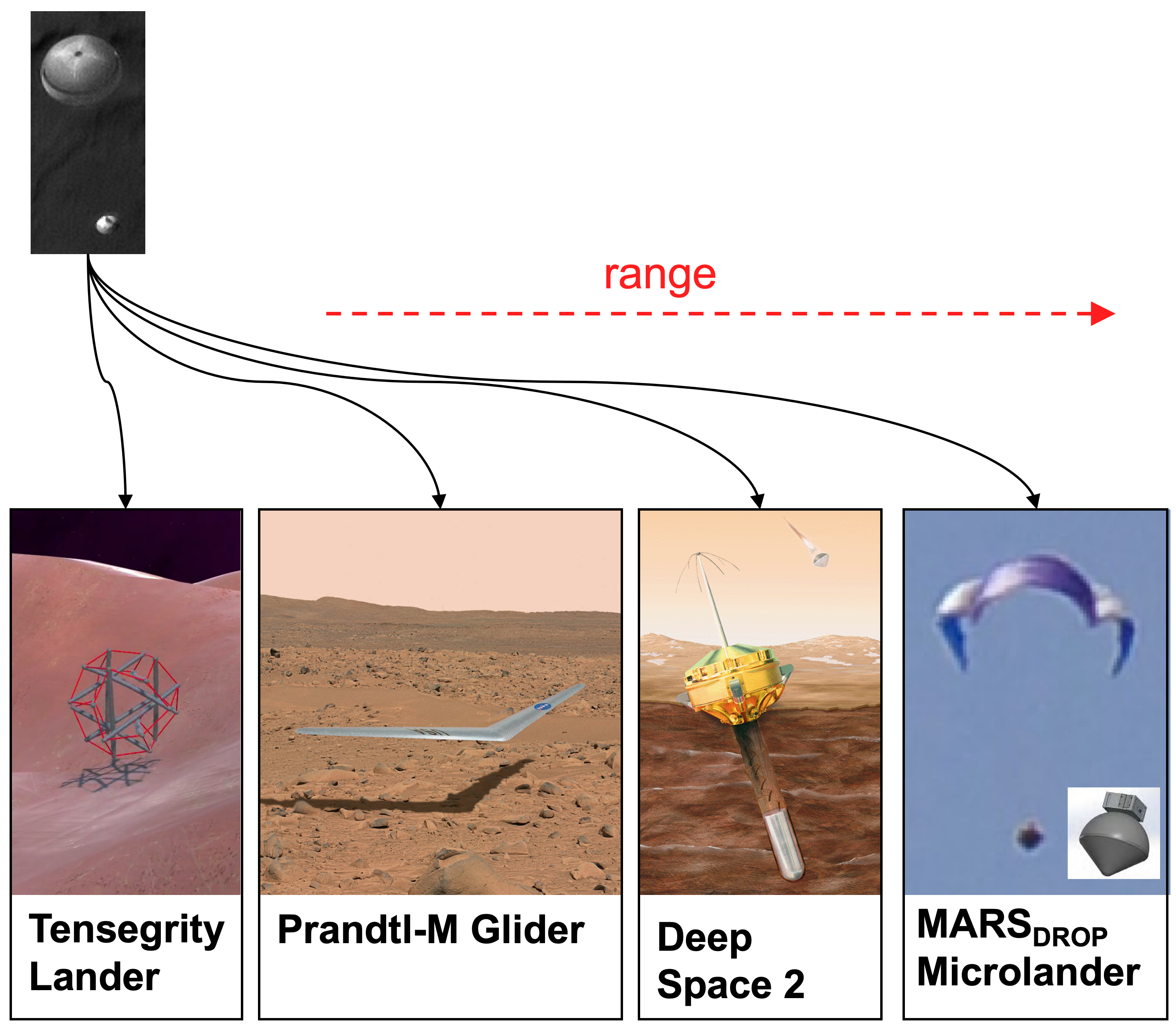}
\caption{Sensor placement assets for sensor deployment on planetary surfaces, with increasing range after deployment.}
\label{fig:technology:sensor-placement:swarm_assets}
\end{figure}

Different science objectives result in different requirements for the spatial distribution of sensors, ranging from global distribution for, e.g., climate science, to accurate regional placement for, e.g., detection of gas plumes. 
Realizing the desired distribution and placement of sensors is a significant technology challenge. A number of technologies  have been developed in the last decade to land and place multiple sensors on planetary surfaces, covering a wide range of values for spatial range, emplacement precision, and cost (Table \ref{tab:technology:sensorplacement}, Figure \ref{fig:technology:sensor-placement:swarm_assets}).

Technologies for deploying sensors on planetary surfaces can be broadly classified into three categories: independent landers, air deployment during descent and landing, and deployment from a single landed asset.

\begin{table}[!h]
  \centering
  \caption{Key properties and performance parameters for selected sensor assets.}
\label{tab:technology:sensorplacement}
  \footnotesize
\begin{tabular}{p{1in}p{1.3in}p{1.5in}p{1.3in}p{1.25in}}
\textbf{Name} & \textbf{Deployment Range} & \textbf{Placement Accuracy} & \textbf{Placement Forces} & \textbf{Payload Size}  \\
\hline
Microlanders (SHIELD) 
& Global %
& $\approx 80 \times \approx 6$ km landing ellipse %
& $< 2000$ Earth g's for $\leq 8$ msec %
& 10 kg (6 kg for science) %
\\
Penetrators 
& Global %
& $\approx 5-10$ km %
& $15000 - 20000$ Earth g's %
& $19$ kg (payload, power, thermal, comm) %
\\
Gliders (Prandtl-M)
& $\approx$30 km, when deployed at $\approx$3.6 km altitude during EDL %
& N/A %
& Impact landing on surface  %
& $605$ g ($230$ g flight system, $375$ g payload and comm) %
\\
Tensegrity Lander 
& N/A %
& N/A %
& Survive a landing speed of $\approx$15 m/sec %
& $5-10$ kg%
\\
\hline
\hline
\textbf{Name} & \textbf{Planetary Bodies} & \textbf{Design life} & \textbf{TRL} & \textbf{Unit cost} \\
\hline
Microlanders (SHIELD) 
& Mars, Venus, Titan (needs atmosphere) %
& 1 year %
& 4 %
& <\$50 million%
\\
Penetrators 
& Moon, Enceladus, comets, Ceres, Mars polar caps %
& 1-2 years %
& 8 %
& \$10M %
\\
Gliders (Prandtl-M)
& Mars, Venus, Titan, Jupiter,  %
& N/A (survives flight) %
& 3%
& $\approx$ \$1M %
\\
Tensegrity Lander 
& Mars, Titan, other rocky bodies%
& N/A%
& 3%
& N/A %
\end{tabular}

\end{table}


\subsubsection{Independent Landers}

Sensors can be deployed to the surface by independent landers, either carried by separate spacecraft or released by a single vehicle upon approach to the body of interest.
Independent microlanders like Mars\textsubscript{DROP} \citep{MarsDrop} and the Small High Impact Energy Landing Device (SHIELD) \citep{Ref:Barba2019_SHIELD},
 and penetrators like Deep~Space~2 \citep{smrekar1999deep} 
are well-suited for establishing a global network of sensors, enabling access to most latitudes and allowing individual sensors to be independently placed across a planetary surface. 

\begin{figure}[!h]
\ifaaspsj
\gridline{
\fig{SHIELD_Lander_concept.png}{0.45\textwidth}{(a) SHIELD Lander concept \citep{Ref:Barba2020_SHIELD}}
\fig{Deep_Space_2_text.png}{0.35\textwidth}{(b) Ballistic penetrator: Deep~Space~2 \citep{smrekar1999deep}}
}
\else
\begin{subfigure}{.5\textwidth}
  \centering
  \includegraphics[width=.8\linewidth]{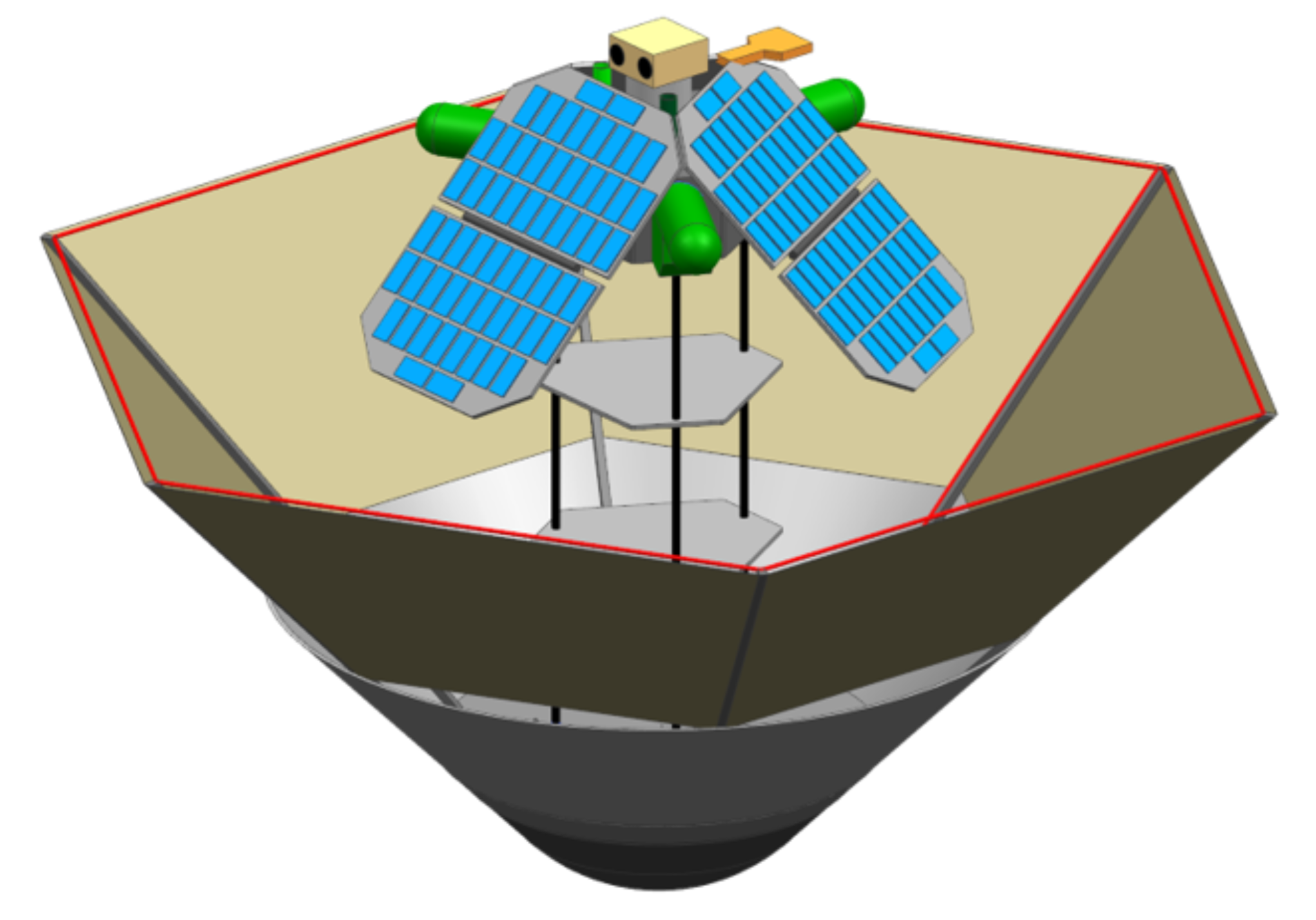}  
  \caption{SHIELD Lander concept \citep{Ref:Barba2020_SHIELD}}
  \label{fig:SHIELD}
\end{subfigure}
\begin{subfigure}{.5\textwidth}
  \centering
  \includegraphics[width=0.8\linewidth]{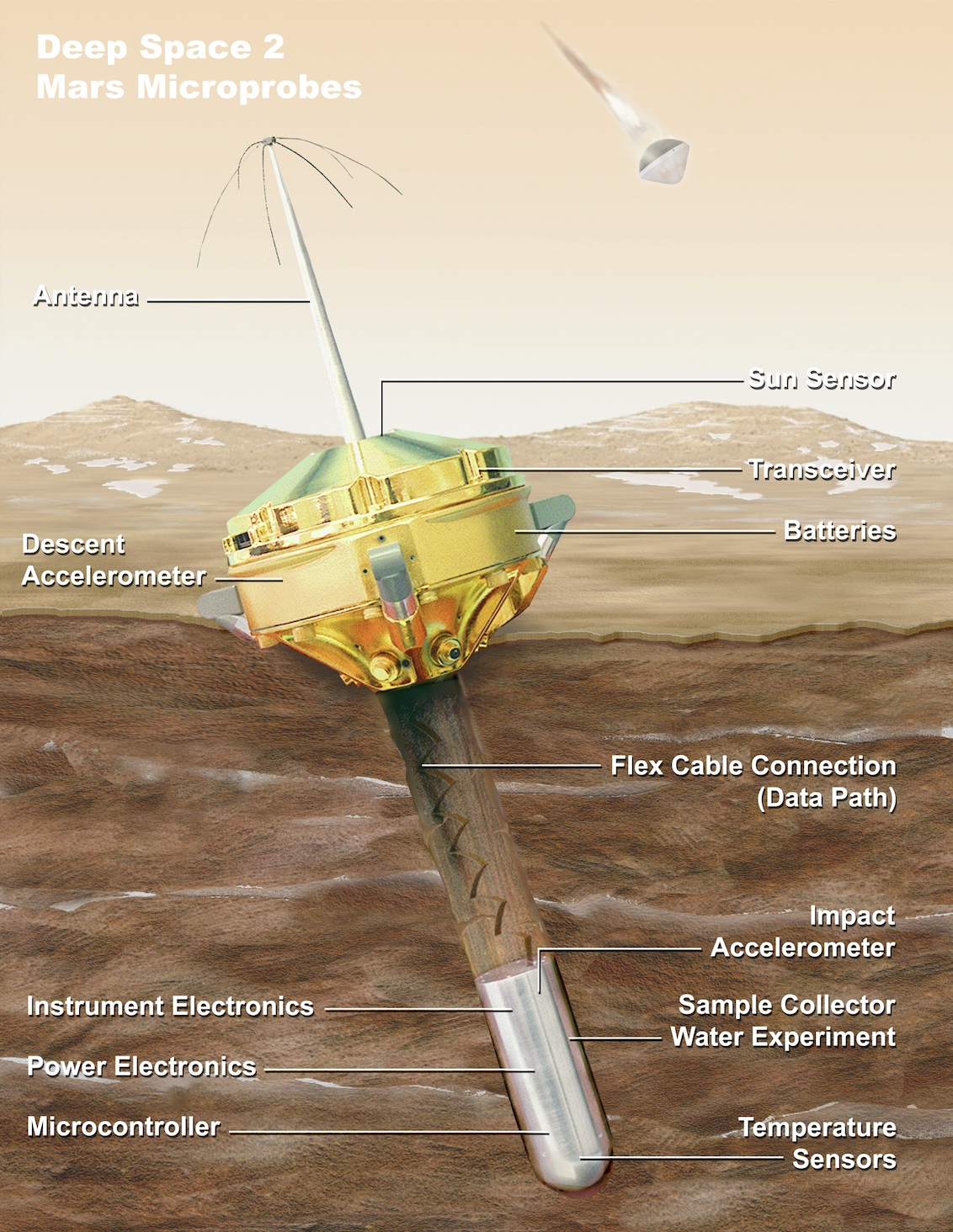}  
  \caption{Ballistic penetrator: Deep~Space~2 \citep{smrekar1999deep}}
  \label{fig:DeepSpace2}
\end{subfigure}

\fi
\caption{Independent Microlanders and Ballistic Penetrators}
\label{fig:Microlanders-and-penetrators}
\end{figure}

Microlanders like SHIELD \citep{Ref:Barba2019_SHIELD} and Mars\textsubscript{DROP} \citep{MarsDrop} (Figure \ref{fig:Microlanders-and-penetrators}(a)) are each fitted with their own EDL package, and travel independently to their landing site.

Penetrators such as Deep~Space~2 [\cite{smrekar1999deep}, Figure \ref{fig:Microlanders-and-penetrators}(b)] 
are typically carried by a single ``carrier'' spacecraft to the planetary target body. Ballistic penetrators detach from the carrier, plummet towards the surface without a  parachute, and bury deep inside the planetary surface. %

\subsubsection{Deployment during Entry, Descent and Landing (EDL) or Approach, Descent and Landing (ADL) Phase \label{sec:deployment-EDL}}

In a second family of approaches, a single ``carrier'' spacecraft protects all sensor assets during the bulk of Entry, Descent, and Landing (EDL) (or, in the case of airless bodies, Approach, Descent, and Landing, or ADL). The sensors are then deployed at much lower altitudes and velocities compared to independent landers. This approach can result in a reduction in the mass and volume allocated to EDL/ADL for individual sensors, increasing the overall science payload capability. However, the lower deployment altitude makes this approach more suited to regional and local sensor distributions, where the maximum distance between sensors is on the order of tens of kilometers \citep{Ref:Bandyopadhyay_IAC19_Prob_Backshell}.

\begin{figure}[!h]
\ifaaspsj
\gridline{
\fig{mars_prandtl-m.jpg}  {.4\textwidth}{(a) Artist's illustration of Prandtl-M glider at Mars. \citep{Ref:PrandtlM_NASA}}
\fig{SUPERball_Ames.png}{0.4\textwidth}{(b) SUPERball v2 robot at NASA Ames \citep{Ref:Vespignani2018_SUPERball}}
}
\else
\begin{subfigure}{.5\textwidth}
  \centering
  \includegraphics[width=0.8\linewidth]{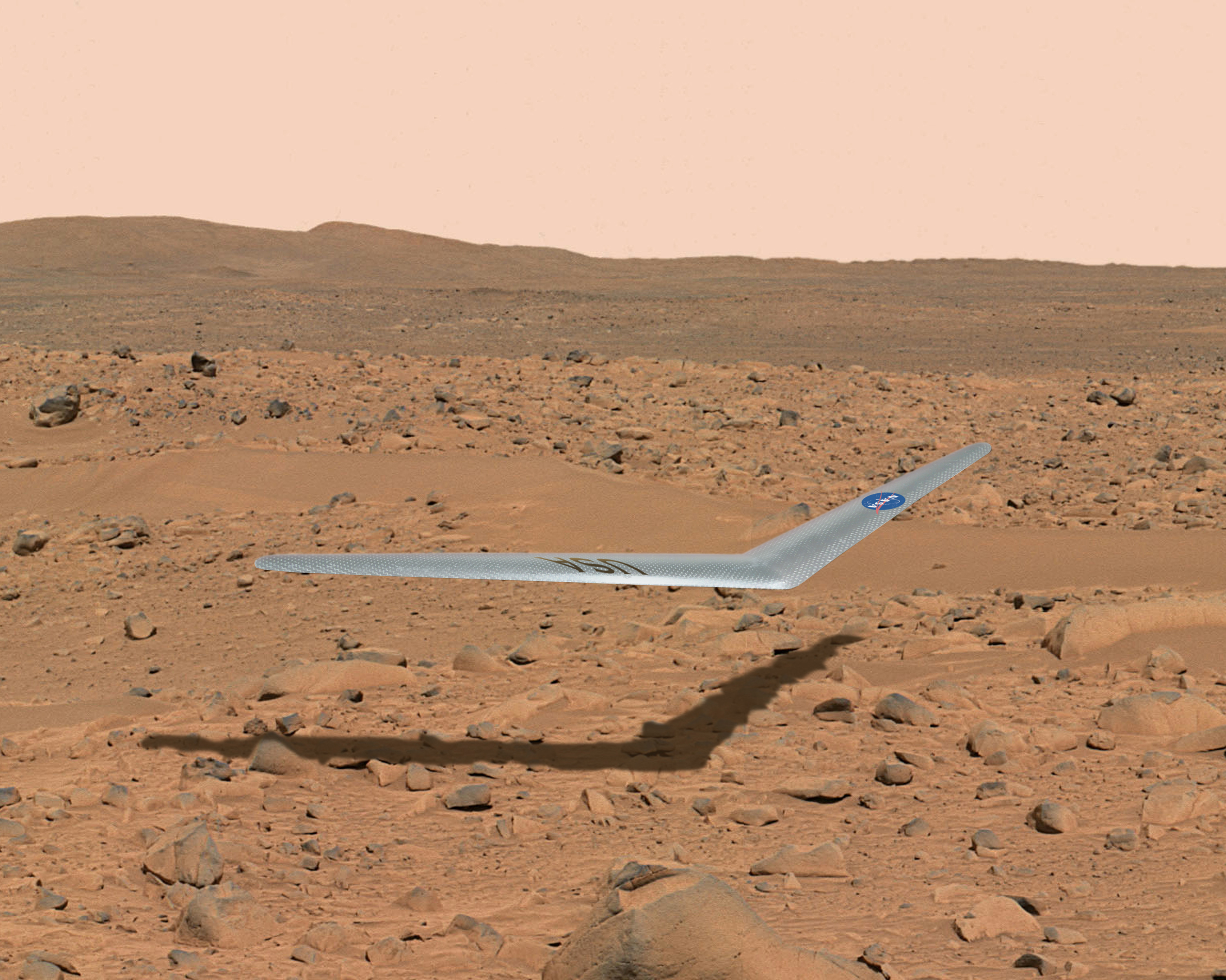}  
  \caption{Artist's illustration of Prandtl-M glider at Mars. \citep{Ref:PrandtlM_NASA}}
  \label{fig:Prandtl_AFRC}
\end{subfigure}
\begin{subfigure}{.5\textwidth}
  \centering
  \includegraphics[width=0.8\linewidth]{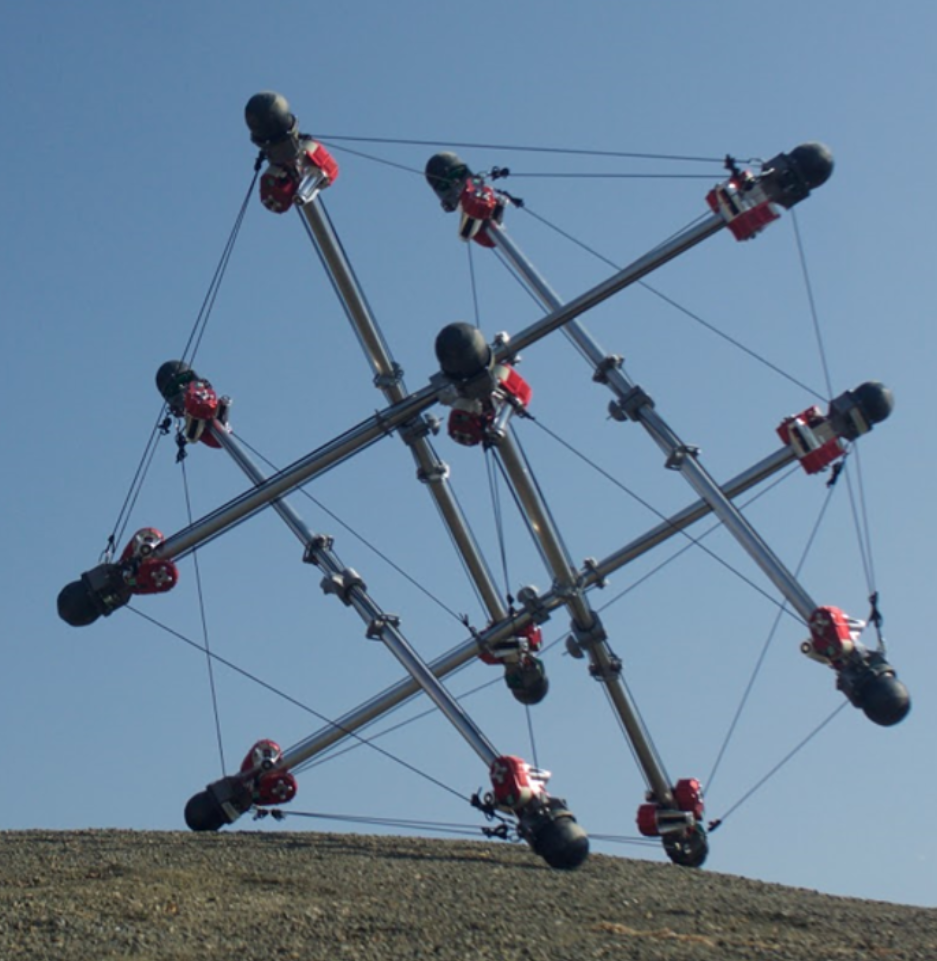}  
  \caption{SUPERball v2 robot at NASA Ames \citep{Ref:Vespignani2018_SUPERball}}
  \label{fig:SUPERball}
\end{subfigure}
\fi
\caption{Gliders and tensegrity landers can be deployed during the final phases of EDL, distributing sensors across regions tens of kilometers across. }
\label{fig:gliders_and_tensegrity}
\end{figure}

Gliders and tensegrity landers/rovers can further increase the deployment range.
For instance, the Prandl-M glider [\cite{Ref:PrandtlM_NASA}, Figure \ref{fig:gliders_and_tensegrity}(a)] is being designed for a Mars piggy-back mission at the NASA Armstrong Flight Research Center (AFRC); deployed at an altitude of $\approx 3.6$ km, it is designed to achieve a range of $\approx 30$ km thanks to a lift-to-drag ratio of $\approx 8$.
Tensegrity landers, deployed during landing, can act as airbag substitutes for the sensors within and also provide mobility once on the surface. Several tensegrity landers have been developed to TRL 4-5, including the SUPERball v2 robot [\cite{Ref:Vespignani2018_SUPERball}, Figure \ref{fig:gliders_and_tensegrity}(b)] and the \emph{sphere-like} tensegrity landers \citep{Ref:Garanger2020soft,Ref:Rimoli2016impact}. Placement forces depend on the payload and the configuration, but in general it would experience loads much lower (1 or 2 orders of magnitude lower) that in any traditional truss-based enclosure.

Deployment during EDL/ADL is especially promising as as an infusion path for distributed instruments as secondary payloads. 
For instance, Martian rover missions typically carry multiple tungsten ballast masses (weighing $\approx$150 kg), which are released during EDL to change the center of gravity of the spacecraft.
A proposed concept \citep{NASABalanceMassChallenge} suggests replacing some of these ballasts with deployable sensor modules which could be distributed over a region spanning tens of kilometers \citep{Ref:Bandyopadhyay_IAC19_Prob_Backshell}.

\subsubsection{Direct Deployment using Rover or Lander} 
Finally, it is possible to also deploy the sensors directly from a rover or lander.
For example, the Mars 2020 mission carried the \textit{Ingenuity}Mars Helicopter, a small rotorcraft that demonstrated the ability for small aerial vehicles to  traverse large distances.
Miniaturized sensors (e.g., \citep{Printable}) could be deployed by a traditional rover or by small, light-weight robotic platforms such as PUFFER/CADRE \citep{PUFFER}.
The key advantages of direct deployment are the high accuracy of sensor placement and the low dynamic load during deployment; however, the spatial extent and distribution of the sensor network is limited by the carrier vehicle's range, which typically precludes large regional or global distributions.

\subsection{On-board Computing}
\label{sec.Computation}

Distributed instruments require on-board computing capabilities to perform data acquisition and analog-to-digital conversion from sensors, signal processing and filtering, including implementing the data synthesis algorithms described in Section \ref{sec.AdaptiveSampling}, localization, synchronization, power and thermal management, and health monitoring, and communications with other sensors and with Earth (possibly through multi-hop protocols such as delay-tolerant networking DTN \citep{burleigh2003dtn}).  

Computing and radiation tolerance requirements will vary significantly depending on the science goals and planetary destination. For instance, a seismology application will require more on-board signal processing capabilities compared to a climate science instrument; and, for the same mission duration, a distributed instrument on Europa will be subject to a harsher radiation environment than a mission to the Martian surface, requiring higher radiation tolerance. 

Space missions beyond Earth's magnetosphere have typically relied on radiation-hardened processors such as BAE systems' RAD750 family and, more recently, the LEON processor developed by the  European Space Agency (ESA). NASA and the Air Force Research Laboratory are also investing in the development of the High Performance Spaceflight Computing (HPSC) chipset \citep{HPSC}, a rad-hard multi-core processor designed to support vision-based applications, model-based reasoning techniques for autonomy, and high-rate data processing for future space missions. 
Recently, there has also been a surge in the adoption of COTS computing solutions for deep space applications, as exemplified by the use of the Qualcomm Snapdragon 801 processor on the Mars Helicopter \citep{MarsHelicopter}. However, these processors and computing platforms have high pin-counts, high cost and, critically, comparatively high power consumption in the tens of watts; this motivates interest in lower-power computing platforms for low-SWaP distributed instruments.

Space-qualified, low-SWaP computing solutions can be classified in two categories: (i) simplified versions of rad-hard processors such as LEON designed for embedded applications, and (ii) low-power COTS microcontrollers that have undergone the qualification process for space applications.

The LEON3FT microcontroller \citep{LEON3FT}, a reduced version of the LEON2-FT processor core developed through funding by ESA, is a good example of the first approach. 
The critical specifications of the LEON3FT are listed in Table \ref{table.COTScomparison}.

A wide range of ultra low-power microcontrollers are available for industrial, automotive and military embedded applications. Thanks to the recent explosion in the number of commercial, low-Earth orbit satellite constellations, many of these components are now also available in space-qualified versions that have undergone radiation testing and provide increased temperature range, reliability and radiation tolerance.
The radiation-hardened version of Texas Instrument's TI-MSP430 microprocessor provides a representative example. The MSP430 consumes 100~$\mu$A/MHz and carries 64~kB of non-volatile ferroelectric RAM (FRAM), which offers higher radiation tolerance compared to SRAM. 

The specifications of some of these radiation tolerant processors and microcontrollers are listed in Table \ref{table.COTScomparison}. Note that all these microcontrollers can be under-clocked to reduce their power consumption, increasing their adaptability for a wider range of applications.

\begin{deluxetable}{cccccccc}
	\tablecaption{Comparing the specifications and radiation tolerance of various COTS computing solutions.}
	\label{table.COTScomparison}
	
\tablehead{
\colhead{Name}& \colhead{\rotatebox{90}{Manufacturer}} & \colhead{\rotatebox{90}{Maximum Clock Frequency (MHz)}} & \colhead{\rotatebox{90}{Total Ionizing Dose (krad)}} & \colhead{\rotatebox{90}{SEU threshold (MeV.cm\textsuperscript{2}/mg)}} & \colhead{\rotatebox{90}{Memory}} & \colhead{\rotatebox{90}{Power}} & \colhead{\rotatebox{90}{Operating Temperature Range ($\degree$C)}}
}

\startdata
LEON3FT &%
ESA, Gaisler &%
50 &%
100 &%
118 &%
192 kB EDAC-protected, &%
300 mW @25 MHz &%
N/A  %
\\
&%
&%
 &%
 &%
 &%
16 MB ROM,  &%
&%
(FPGA-%
\\
 &%
&%
 &%
 &%
 &%
256 MB SRAM &%
&%
dependent)%
\\
MSP430FR5969-SP &%
Texas Instruments &%
16 &%
50 &%
85&%
64 kB FRAM &%
 330 $\mu$ W/MHz active&%
-55 -- 105 %
\\
SAMV71Q21RT &%
Microchip  &%
300 &%
30 &%
60 &%
 2 MB Flash,  &%
15~mW @ 12 MHz, &%
-55 -- 125%
\\
 &%
  &%
 &%
 &%
 &%
384 kB SRAM &%
 300~mW @ 300 MHz&%
\\
ATmegaS128 &%
Microchip  &%
8 &%
30 &%
62.5&%
128kB Flash, 4kB SRAM&%
20~mW @ 10~MHz &%
-55 -- 125%
\\
\enddata
\end{deluxetable}

Another alternative is to use commercially available, radiation-tolerant field programmable gate arrays (FPGAs). While FPGAs typically have higher power consumption compared to microprocessors, they can be easily optimized for the specific data processing needed for a given science application. Xilinx provides rad-hard FPGA solutions that can tolerate 300~krad of TID and have single event latch-up threshold of 100~MeV.cm\textsuperscript{2}/mg. Their products are specifically advertised for interplanetary and deep-space applications. %

\subsection{Power for Distributed Sensors}
\label{sec:technology:power}

\begin{figure}[h]
\centering
\includegraphics[width=.7\textwidth]{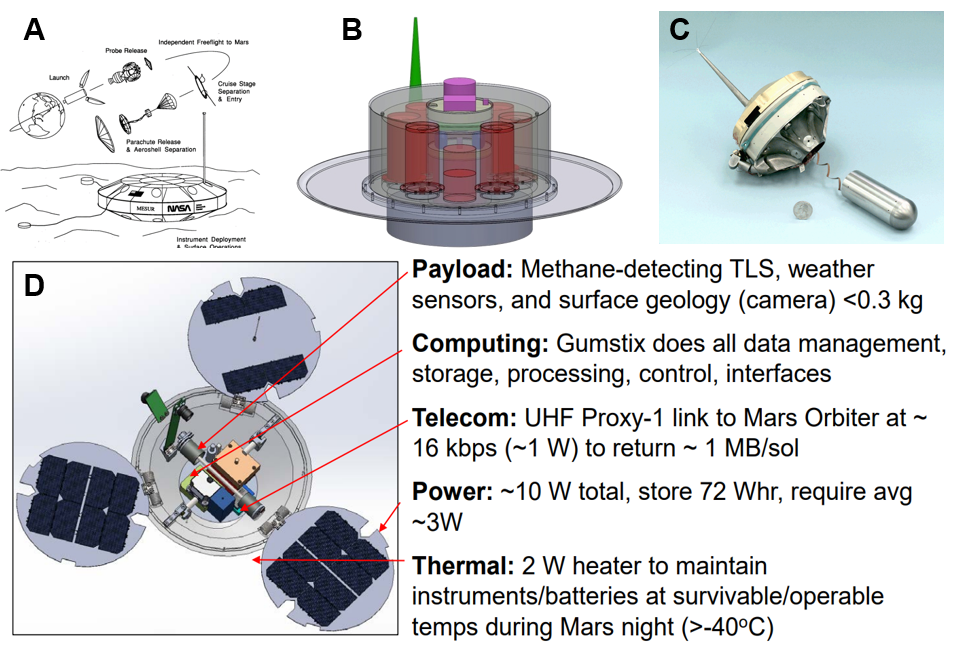}
\caption{ (A) MESUR concept, (B) MASER concept, (C) DS-2 Mars Microprobe, and (D) Mars\textsubscript{DROP} concept.}
\label{fig:technology:power:concepts}
\end{figure}

Individual sensor nodes within a distributed instrument will require a power source capable of operating on the planetary surface of interest.  The required power levels will vary depending on the choice of processor, radio, instruments, and required heating and duty cycle of the electronics. Choice of a specific power source will depend on mission duration, distance from the Sun, expected temperature levels, radiation levels, and requirements related to impact resistance. 
Several small Mars missions or concepts have been developed over the years, which can serve as reference missions to highlight different power source approaches for distributed sensor networks (Fig. \ref{fig:technology:power:concepts}). The MESUR concept \citep{hubbard1992mars} proposed use of a small radioisotope thermoelectric generator (RTG); The DS-2 microprobes  \citep{smrekar1999deep} were powered by low-temperature-capable primary batteries only; the Meteorology and Seismology Enabled by Radioisotopes (MASER) concept \citep{lorenz2014maser} was baselined using a small RTG, employing a radioisotope heating unit (RHU) heat source along with supercapacitors to support peak power requirements; and the Mars\textsubscript{DROP} concept \citep{MarsDrop} proposed the use of small solar arrays, in combination with lithium-ion batteries for energy storage.
A detailed summary of options is provided in the sections below.

\subsubsection{Primary Batteries}

\begin{figure}[h]
\centering
\includegraphics[width=.5\textwidth]{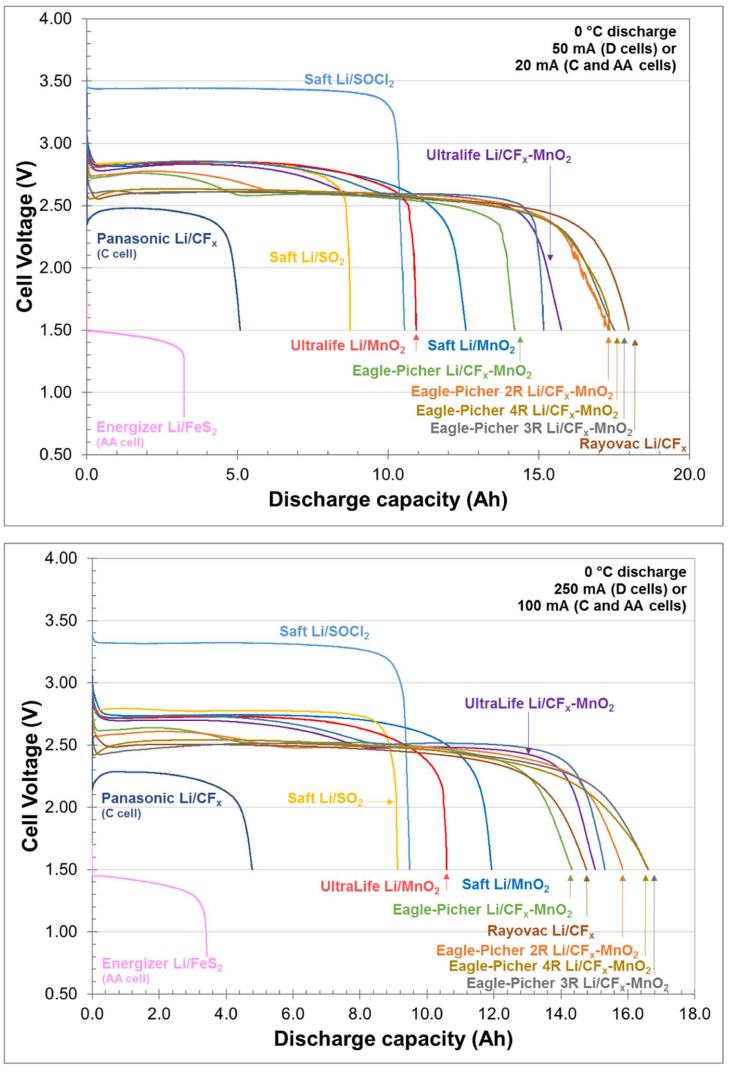}
\caption{Representative primary batteries at 0$^\circ$ C. From \cite{krause2018high}.}
\label{fig:technology:power:primary-batteries}
\end{figure}

Primary batteries are capable of providing electrical power to various loads, without the ability to recharge. Since they are not rechargeable, a wider range of electrode materials can be chosen relative to state-of-the-art secondary technologies, such as lithium-ion chemistries. All sensors can, in principle, be powered by high specific energy, non-rechargeable primary batteries, with the mission duration dictated by the specific energy of the batteries, the mass allocated for these batteries, and the average power of the loads. They are typically considered for missions lasting days to weeks, although longer missions with extended low power/quiescent periods could be considered, particularly with the availability of new high specific energy chemistries. Recent capacity vs. voltage data are shown for a range of battery types and chemistries, in Fig. \ref{fig:technology:power:primary-batteries} \citep{krause2018high}.

The high specific energy Li/CF\textsubscript{x} chemistry in a D-cell format seen in Fig. \ref{fig:technology:power:primary-batteries} is a unique design with no flight heritage. This cell, which was developed by the Europa Lander Pre-Project \cite{Hand_2022} offers a very high cell-level specific energy (>700 Wh/kg), and is ideal for low rate applications. D-sized cells now have a capacity approaching 20 Ah and a mass of $\approx$ 70 g each, and provide 2.6 V continuously under load. These cells typically deliver approximately 50 Wh of total energy each. Therefore, two cells in series can provide 1 W average electrical power at 5 V for $\approx$ 100 hours.  In addition, these cells can produce a nearly constant amount of thermal energy during discharge, depending on the discharge rate, meaning $\approx$ 1 W\textsubscript{th} is available for thermal management. One concern is the voltage delay, or drop in voltage, which is present at start-up, in particular, when operating at low temperatures. However, with proper thermal design, the heat generated during discharge can be retained in the battery pack to keep the cells sufficiently warm. In fact, heat generation is such an issue that terrestrial packs typically use a modified form, Li/CF\textsubscript{x}-MnO\textsubscript{2}, to reduce the amount of heat generation. 
Alternatively, if inherently low temperature operation ($-40^\circ$ C and below) is required and higher rate capabilities needed (>1 A), a liquid cathode chemistry such as Li/SO\textsubscript{2} or Li/ SOCl\textsubscript{2} can be used \citep{bugga2020energy}. These cell types have significant flight heritage. In fact, a modified version of the Li/SOCl\textsubscript{2} chemistry was developed for the DS-2 mission, for very low temperature operations. For high temperature Venus probes, batteries capable of operating at Venus surface temperatures ($+465^\circ$C) are being developed under the NASA HOTTech program \citep{bugga2020new}.

\subsubsection{Solar Cells}

Several solar cell options exist for surface missions on the Moon and on Mars. Although solar cells can be used in larger orbiters and fly-by missions to Jupiter and even to the Ice Giants, use of solar cells for power generation \revmm{in landers (and in particular in distributed instruments comprised of small sensors)} is practically limited to the surface of Mars due to the much lower solar flux at Jupiter distances ($\leq 50$ W/m$^2$ depending on latitude and season vs. 1366 W/m$^2$ at Earth) and beyond. Spacecraft solar arrays are typically very large at these distance, in order to produce appreciable power levels. They would not be practical, for example, for small distributed sensors on planetary surfaces/moons such as Europa, Enceladus or Titan (with the latter's atmospheric opacity introducing a further challenge to use of solar cells). State-of-the-art cells are represented by those on the Mars Helicopter, using inverted metamorphic multi-junction (IMM) type cells with an efficiency approaching 35\% (beginning-of-life, $28^\circ$C and under the standard spectrum outside the Earth’s atmosphere or Air Mass 0 [AM0] conditions) \citep{sharps2017next}. Accumulation of dust degrades power output in the Mars environment. Due to the extreme temperatures and need for robust thermal management, application in lunar environments for small, long-duration probes is unlikely.

\subsubsection{Radioisotope Power Sources}

\begin{figure}[h]
\centering
\includegraphics[width=.7\textwidth]{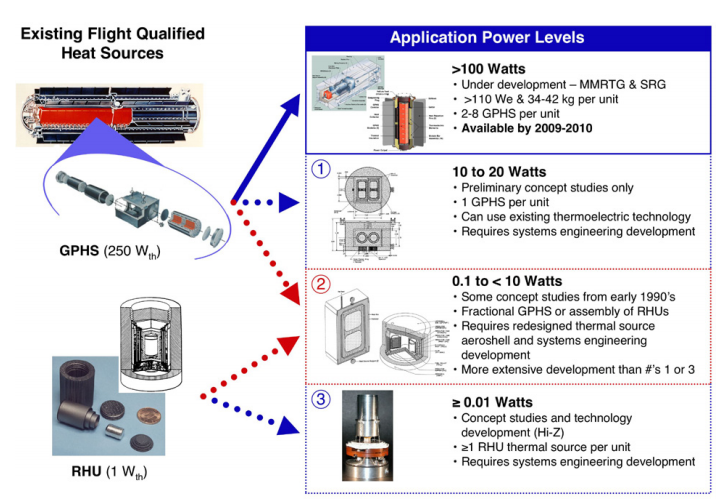}
\caption{Range of radioisotope thermoelectric generators, producing power from about 40 mW to >100 W. From \cite{schmidt2005benefit}.}
\label{fig:technology:power:rtgs-range}
\end{figure}

As seen with the proposed MESUR and MASER probes on Mars, small RTGs can be considered for use as both a heat and electrical power source. The RTG uses an arrangement of power-generating thermocouples, which generate a voltage across the temperature difference between the couple and space. The heat is provided by small amount of radioisotope “fuel” (typically Pu-238, although alternative fuels such as Sr-90 are under consideration). To fit in such small spaces, a fractional General Purpose Heat Source (GPHS) can be used, which produces $\approx$ 250 W of thermal energy from the decay of Pu-238 at the time of manufacture. Alternatively, the standard 1 W\textsubscript{th} RHU can be used to generate mWs of power. These concepts are depicted in Fig. \ref{fig:technology:power:rtgs-range} \citep{schmidt2005benefit}.

\begin{figure}[p]
 \centering
\includegraphics[width=\textwidth]{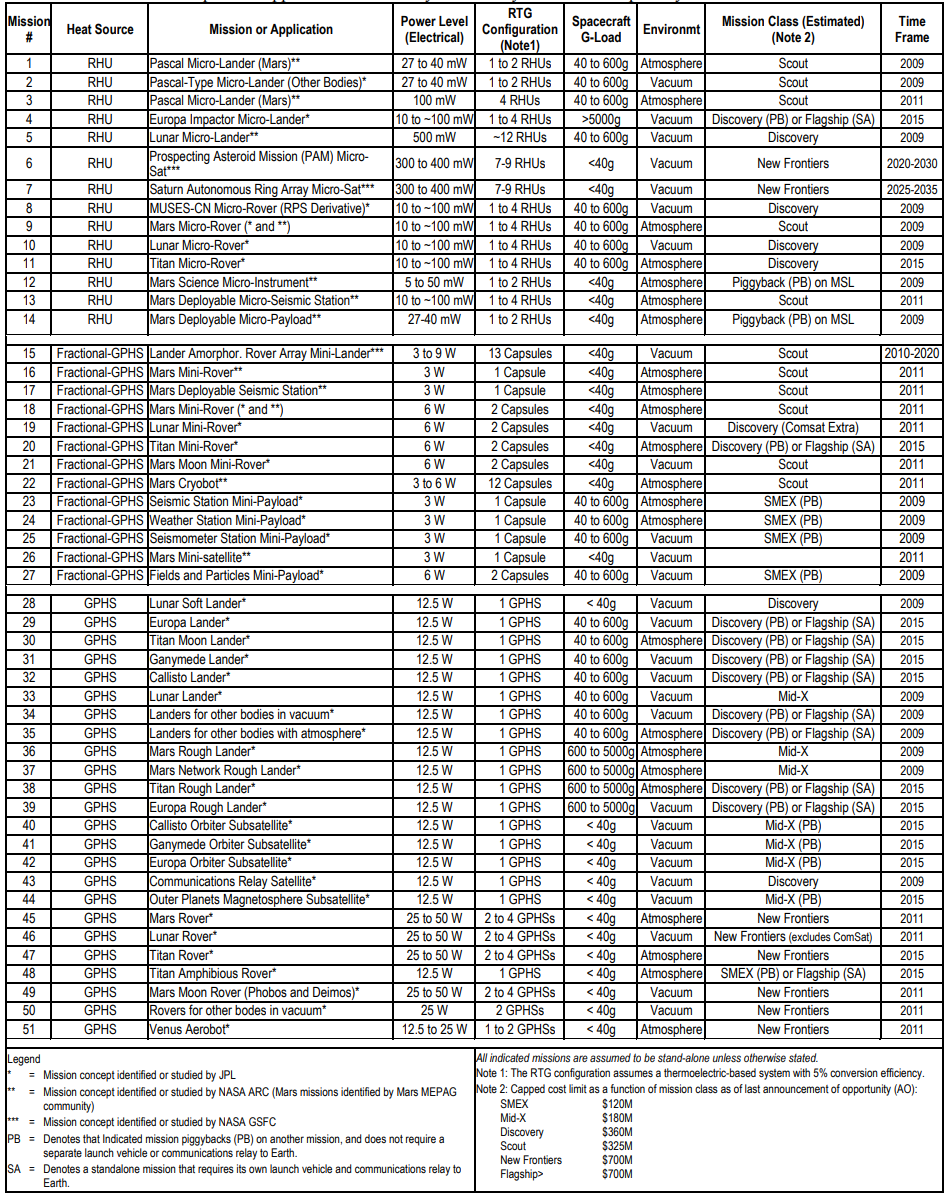}
\caption{Missions concepts enabled by small RTG power sources. From \cite{schmidt2005benefit}.}
\label{fig:technology:power:rtg-concepts}
\end{figure}

A wide range of missions could be enabled by these small RTG systems. A summary of past studies is given in Fig. \ref{fig:technology:power:rtg-concepts}, below. RTGs in the <50 mW to 25 W range can be envisioned, using an RHU heat source (27-500 mW), a fractional GPHS (3-9 W) or a full GPHS (12.5-25 W). Another potential technology for consideration are beta-voltaic devices, which use a \emph{p-n} junction to generate a voltage when exposed to a beta-emitter such as tritium \citep{revankar2014advances}; however, to date, these devices have only demonstrated $<\mu$W power levels.
 
\subsubsection{Energy Storage}

To support peak power requirements, or to provide power during nighttime for solar-powered missions, some form of energy storage is needed. Typical space-rated options include lithium-ion batteries and supercapacitors \citep{bugga2020energy}. Small cylindrical lithium-ion cells suitable for use in small probes and manufactured by companies such as Sony, Panasonic and LG Chem have been used for the last 20 years in space, going back to the first use in the PROBA small satellite mission (European Space Agency). These cells typically require no cell balancing electronics, and offer 3.5 Ah of capacity at up to 4.1 V. Cells are typically rated for use to about -20$^\circ$C, although cells customized with low temperature electrolytes can operate to -40$^\circ$C or below.
Supercapacitors rely on energy storage at the electrochemical double-layer between an electrolyte and a high surface area carbon material. These devices are also known as electrical double-layer capacitors (or EDLCs). The advantages of this technology include very high specific power (>10 kW/kg), very long cycle life (10$^6$ cycles) and very wide temperature operation (typically rated between -40 and 60$^\circ$C). The high specific power results from the very low internal resistance of these devices. Testing of actual devices at JPL has indicated EDLCs can operate to -50$^\circ$C, and prototype devices with modified electrolytes have been constructed and tested for operation down to -80$^\circ$C. The major disadvantage is the low specific energy (5 Wh/kg) relative to traditional battery technologies. They are ideal when short bursts of energy are needed, for example, in communications. 
A very low power draw sensor with periodic high pulse conditions would be an ideal use case for pairing a supercapacitor with a long-life source such as a small RTG. This configuration could support operations to very low temperatures with reduced thermal management requirements, and would feature a very long life (with no concerns over life-limiting mechanisms associated with lithium plating in lithium-ion batteries charged at low temperatures). For applications where greater energy is required, hybrid battery/supercapacitor modules may still be advantageous \citep{bugga2020energy}. This allows the battery to provide a constant source of energy, with the supercapacitor providing pulse power. This allows the battery to be sized smaller, with the battery being buffered from any high current pulses (thus extending life time of the energy storage element).

\subsubsection{System Considerations}

Apart from a limited subset of missions to the Martian surface, future deep space sensor networks are likely to require some type of primary battery power source, or, alternatively, a radioisotope-based power source producing power in the <1 to 10 W range. The MASER mission concept study offers a scoping of requirements for a power system for a small distributed sensor node. The instrument package for this Mars probe featured pressure/temperature sensors (2.6 mW) generating 10 bps of data and a seismometer (65 mW) generating 30 bps. These values included margin, and assumed that the sensors would be continuously on. In addition, an optical monitor (26 mW) generating 1.6 bps data and a wind sensor (325 mW) generating 24 bps were included, both operating at an approximate duty cycle of 8\%. The telecommunication capability required by the data generated was assumed to be about 5 Mbit/day. The transmitter was assumed to require 3.3 W, at a duty cycle of 1.4\%, with the electronics requiring 65 mW of continuous electrical power. This resulted in a power draw of approximately 200 mW continuous, and an energy requirement of about 4 Wh/day.

\begin{figure}[h]
\centering
\includegraphics[width=.7\textwidth]{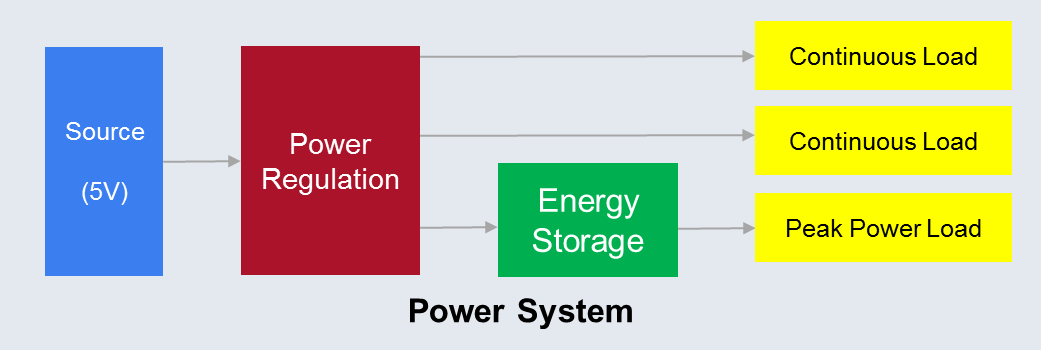}
\caption{Simplified distributed sensor network power system block diagram.}
\label{fig:technology:power:block-diagram}
\end{figure}

The power bus was assumed to operate at 5 V. The design assumed the use of six 40 mW\textsubscript{e} RHU RTGs, generating a total of 240 mW\textsubscript{e}. The balance of 1 W thermal from each RHU could be used for thermal management (along with use of aerogel insulation assumed at 0.016 W/m.K). To provide the peak power needed for data transmission, four 2.7 V, 650 F supercapacitor cells were included, with two in series to support the 5 V bus and two more in parallel to increase capacity of the energy storage element. A generalized block diagram for a small power source is depicted in Fig. \ref{fig:technology:power:block-diagram}, along with a summary of power source options (Table \ref{tab:technologies:power:summary}).

\begin{table}[h]
\caption{Summary of Representative Power Source Technologies for Distributed Sensor Systems}
\label{tab:technologies:power:summary}
\footnotesize
\begin{tabularx}{\textwidth}{cXXXX}
\hline
\hline
\textbf{Power Source} & \textbf{Technology} & \textbf{Specifications} & \textbf{Performance} & \textbf{Comments} \\
\hline
Primary Battery & Li/SO$_2$ (Saft) & D-cell, 85 g, T=-60 to 85$^\circ$C & 7.5 Ah, 3.0 V, $\approx$ 300 Wh/kg & Low temperature performance; flight qualified \\
Primary Battery & Li/CF$_\text{x}$ (EaglePicher) & D-cell, 70 g, T=0 to 100$^\circ$C & 20 Ah, 2.6 V, $\approx$ 700 Wh/kg & Generates $\approx$ 1 W thermal power for every 1 W$_{e}$; (developed for the Europa Lander concept mission)\\
Solar Array & ZTJ 3\textsuperscript{rd} Generation Triple-Junction Solar Cell and IMM LILT Solar Cell (SolAero) & Coverglass Interconnected Cell available for integration onto solar panels & 29.5\% efficiency (maximum power point),
V$_\text{oc}$=2.726 V,
J$_\text{sc}$=17.4 mA/cm$^2$ (AM0),
84 mg/cm$^2$ cell mass,
(35\% for IMM) & Used for InSight, M2020, Mars Helicopter \\
Small RTG & RHU-RTG  (Hi-Z) & Bi-Te thermocouples with RHU heat source & 35 mW at 5 V per unit & Uses 1.1 W\textsubscript{th} radioisotope heater unit\\
Beta-voltaic & Tritium Hydride Thin Film Device
(City Labs) & T=-50 to 150$^\circ$C & V$_\text{oc}$=0.77 V, I$_\text{sc}$=500 nA & Current product line\\
\hline
\end{tabularx}
\end{table}

\afterpage{\clearpage}

\subsection{Thermal Control}

The thermal design of a distributed instrument depends heavily on the target body. Ambient temperatures and the presence or absence of an atmosphere have a major impact on the solutions that need to be implemented at each sensor node. Broadly speaking, two classes of thermal mitigation strategies can be adopted, namely, passive and active strategies. 

\subsubsection{Passive Solutions}
Passive solutions involve the use of:
\begin{enumerate}
	\item Thermal coatings to change the emissivity of the radiative surfaces;
	\item Manufacturing materials chosen smartly, keeping in mind their thermal properties;
	\item Multi-layer insulation (MLI) blankets for thermal isolation. In the presence of an atmosphere, the layers can be thermally shorted; in this case, hydrogel-based solutions can be used for thermal isolation;
	\item Heat straps for effective transfer of heat across components.
\end{enumerate}

Other solutions such as heat pipes don't tend to work very well in the presence of gravity, and are less relevant at the length scales typical of miniaturized sensor packages. Batteries tend to be the most thermally sensitive sub-system in many applications; hence, the goal is often to locate the battery so as to maximize the transfer of heat generated by the electronics to the battery module. Additional thermal constraints could be imposed by the specific sensors and their operational temperature range.

\subsubsection{Active Solutions}
As far as active solutions are concerned, heating pads and coils can be employed in the design of the system. This essentially turns the thermal problem into a power problem: the question of how long the temperature of the system can be kept within the desired range becomes a question about how much electrical power is available. One solution for a resource-constrained system is to have short bursts of activity when the module is brought up to its operating temperature, and then put into a cold sleep mode for the remaining period to save energy resources. 

RHUs that contain a small amount of Pu-238 as a source of heat have been used on space missions in the past; however, these tend to be expensive and are challenging in terms of planetary protection, safety concerns and other procedural burdens that they impose on the system. 

To summarize, the challenges associated with the thermal management on distributed instruments can be solved using a combination of existing mature techniques and by designing a smart, resource-aware concept of operations. The thermal management solutions devised for the Mars Helicopter \citep{MarsHelicopterThermal} serve as a good example of how thermal control can be achieved on miniaturized systems in a small form factor.

\subsection{Fit with Science Concepts and Technology Gaps}
\label{sec:technology:gaps}

We are now in a position to assess whether state-of-the-art technologies are sufficiently mature to support the science concepts in Section \ref{ch:sciencecase}, and identify technology gaps.

\paragraph{Localization and Synchronization} The high-priority science concepts discussed in Section \ref{ch:sciencecase} only require comparatively modest localization requirements, which can be readily fulfilled with state-of-the-art orbital localization techniques. Seismology requires accurate clock synchronization, but a combination of RF synchronization (aided by an orbiter) and high-precision on-board clocks (e.g., a chip-scale atomic clock) hold promise to provide extremely accurate synchronization. For other concepts, low-power, on-board oscillators and occasional synchronization with an orbiter are likely to deliver sufficient synchronization accuracy.

\paragraph{Communications} Distributed instruments for global climate investigations, trace gas detection, and magnetometry will generate modest data volumes. In contrast, data rates for seismology distributed instruments are extremely high, and likely to be higher than what can be relayed to Earth from planetary surfaces \revmm{beyond the Moon} with existing communication technology; for instance, InSight was unable to downlink all the data collected by even a single seismometer from Mars. On-board data summarization techniques can partially address this problem by compressing data to only capture relevant events, and are therefore highly attractive for seismic distributed instruments.

The regional and global scale of the proposed distributed instruments makes \emph{direct} sensor-to-sensor communication highly challenging; therefore, any sensor-to-sensor communication is likely to have to be mediated by an orbiter, or relayed through Earth. For landed assets, a small SWaP is desirable for EDL reasons. For this reason, an orbiter is an especially attractive solution to relay data to Earth, since the use of an orbiter relay removes the need for a high-gain antenna on individual landed assets. %

The science concepts discussed in Section \ref{ch:sciencecase} do not require extreme hardware miniaturization: the number of required sensing units is four to sixteen, and the SWaP of the proposed sensors, alone, is in the range of tens of grams to a few kilograms. Accordingly, communication transponders and low-gain antennas developed both at NASA and in the private sector, which offer state-of-the-art capabilities in a 1/4U to 1U package, are likely to be adequate for surface-to-orbiter communications.
For what concerns communication protocols, DTN is a proven technology \revmm{that can provide multi-hop communication over time-varying RF links, supporting data relay and instrument autonomy. }

\paragraph{Data Synthesis, Spatial Estimation, and Adaptive Sampling} On-board instrument autonomy is critical for data reduction in seismology applications, where sensing nodes will produce data volumes that are likely to exceed available bandwidth to Earth. In addition, on-board autonomy could allow distributed instruments for atmospheric science and trace gas detection to implement reactive science paradigm where the DI is quiescent until an event of interest is observed by a sensor node, resulting in significantly reduced power usage. While algorithms for knowledge compression, data prioritization, and reactive science exist, further maturation is required before these can be incorporated in distributed instruments.

\paragraph{EDL and Sensor Placement} Technologies for delivery and placement of sensing units are highly sensitive to the destination body. This contrasts with other technology areas such as localization, synchronization, communications, computing, and autonomy, which are less sensitive to the target planetary surface. 

The high-priority science concepts under consideration all require regional-to-global coverage;
accordingly, EDL and sensor placement solutions based on independent landers appear most suitable, whereas deployment during EDL and direct deployment from vehicles are less applicable (with the possible exception of trace gas detection). The selection of a specific lander asset is unique to the sensor suite and planetary surface of interest; our analysis suggests that microlanders and penetrators hold promise to satisfy the requirements of distributed instrument under consideration, while offering higher technology maturity compared to gliders and tensegrity structures. None of the science concepts under consideration strictly requires local mobility after landing to achieve their primary science goals.

While a number of technologies to deliver multiple small payloads on planetary surfaces are under development, all existing approaches except for penetrators (which impose steep mechanical constraints on instruments) are at comparatively low TRL. Accordingly, additional technology development focused on specific bodies of interest is required before distributed instruments can be realized; efforts such as SHIELD \citep{Ref:Barba2019_SHIELD} are well-aligned with this goal.

\paragraph{On-board Computing} For atmospheric science, trace gas detection, and magnetometry, the need for on-board computing is comparatively limited: a straightforward mission concept might simply task the distributed instrument with collecting and forwarding readings. Accordingly, existing COTS rad-hard, low-power computing units are likely adequate. For applications requiring on-board autonomy, e.g., seismology, existing computing solutions such as LEON are also highly likely to be adequate, at the price of increased power demand.

\paragraph{Power for Distributed Instruments} A limited subset of concepts requiring regional deployment on the Moon or Mars at low latitudes (e.g., a regional trace gas distributed instrument on Mars, or a seismic network on the Moon) may be able to use solar panels \revmm{and rechargeable secondary batteries} for long-term operations. In contrast, most distributed instruments for planetary surface science will require some type of primary battery power source or, alternatively, one using a radioisotope-based power source producing power in the <1 to 10 W range. Even in favorable environments such as the Martian surface, RTG sources or large primary batteries will be critical to enable survival at high latitudes, which is necessary where the underlying science question requires global data collection. High specific energy primary battery cells based on the Li/CFx chemistry and in a D-sized cell format with a specific energy >700 Wh/kg have been developed as part of the Europa Lander concept development project. Small RTG power sources are under active development, but significant additional investment is required to bring them to maturity.

\paragraph{Thermal Control} Thermal control for small landed packages is a relatively well-understood problem: the thermal management solutions devised for the Mars Helicopter \citep{MarsHelicopterThermal} can serve as a good reference for how thermal control can be achieved on miniaturized systems. However, thermal control is likely to be a key driver of power and energy requirements, further emphasizing the need for technology advances in that area.

In summary, additional research and development in (i) EDL and sensor placement solutions, (ii) instrument autonomy (in particular for seismology applications), and (iii) power solutions are required to bring distributed instruments to reality, whereas all other technology areas are comparatively mature.

Figure \ref{tab:enabling-technologies} summarizes these conclusions, classifying the suitability of state-of-the-art technologies to support each of the science concepts presented in Section \ref{ch:sciencecase}. %

\begin{figure}[!ht]
\centering{}%
\begin{tabular}{r|cccc}
\multicolumn{1}{r}{\textbf{Technology}} &  
\rlap{\rotatebox{45}{Magnetometry for Internal Composition}} & 
\rlap{\rotatebox{45}{Global Climate and Weather}} & 
\rlap{\rotatebox{45}{Seismology}} & 
\rlap{\rotatebox{45}{Trace Gas Detection}}
\\
\midrule
Localization & \coloredpie{360}{blue} & \coloredpie{360}{blue}  & \coloredpie{360}{blue} & \coloredpie{360}{blue}  \\
Synchronization & \coloredpie{360}{blue} & \coloredpie{360}{blue}& \coloredpie{360}{blue} & \coloredpie{360}{blue} \\
Communications & \coloredpie{360}{blue} & \coloredpie{360}{blue}& \coloredpie{360}{blue} & \coloredpie{360}{blue} \\
Instrument Autonomy & \coloredpie{270}{green} & \coloredpie{270}{green} & \coloredpie{180}{orange}  & \coloredpie{270}{green}  \\
EDL and Sensor Placement & \coloredpie{180}{orange} &\coloredpie{180}{orange} &\coloredpie{180}{orange} &\coloredpie{180}{orange}  \\
Computation and Storage & \coloredpie{360}{blue} & \coloredpie{360}{blue}  & \coloredpie{270}{green} & \coloredpie{360}{blue}  \\
Power & \coloredpie{180}{orange} & \coloredpie{180}{orange}& \coloredpie{180}{orange}&  \coloredpie{180}{orange} \\
Thermal & \coloredpie{360}{blue} & \coloredpie{360}{blue}  & \coloredpie{360}{blue}  & \coloredpie{360}{blue}  \\
\bottomrule
\end{tabular}
\begin{tablenotes}
\item[1] \coloredpie{360}{blue} : the technology is sufficiently mature for infusion in flight missions (7--9 TRL)
\item[2] \coloredpie{270}{green} : the technology is currently under development, but quite mature (5--6 TRL)
\item[3] \coloredpie{180}{orange} : the technology is currently under development, but not very mature (3--4 TRL)
\end{tablenotes}
\caption{Maturity of enabling technologies for each of the science concepts presented in Section \ref{ch:sciencecase}.}
\label{tab:enabling-technologies}
\end{figure}

\section{Concluding Remarks and Recommendations}
\label{ch:conclusions}

The goal of this paper is to assess the science promise and technology feasibility of distributed instruments, and, specifically, to
(i) survey high-priority scientific questions that can be uniquely addressed by distributed instruments with existing or tipping-point technologies; and, (ii) identify and survey technologies that should be further developed to ensure the technology feasibility and to maximize the scientific returns of promising distributed instruments;

We are now in a position to report our findings.

\subsection{High-Priority Science Questions}

\begin{displayquote}
\emph{Distributed instruments are uniquely well positioned to address a number of high priority scientific questions.}
\end{displayquote}

Our survey unequivocally shows that distributed instruments hold unique promise to address high-priority questions that cannot be addressed with either remote sensing from orbiters, or point measurements from individual landers or rovers.

We identified four broad scientific areas that can \emph{only} be investigated through geographically distributed, spatially and temporally correlated measurements --- i.e., the type of measurements that can only be provided by a distributed instrument, and which are clearly identified as high priority in Decadal Surveys, and could feasibly be addressed with existing and tipping-point technologies:
\begin{itemize}
\item \textbf{global climate and weather} on Mars;
\item \textbf{localization of seismic events} on rocky and icy bodies;
\item \textbf{magnetometry for internal composition} across the Solar System; and
\item \textbf{trace gas detection}, with a focus on Mars.
\end{itemize}

For each of these topics, we identified its scientific importance according to the most recent Decadal Survey, drafted a high-level distributed instrument architecture, identified critical technological challenges that should be addressed, and laid out a preliminary strategy for overcoming these challenges. 

We remark that, in our search, we focused exclusively on the scientific \emph{need} for distributed instruments, i.e., on questions that \revmm{require geographically distributed, spatially and temporally correlated observations.} 
Distributed instruments could also offer additional \emph{engineering benefits} (e.g., faster coverage, lower cost, or higher resilience) compared to monolithic instrument for other science questions. The investigation of such engineering benefits is beyond the scope of this paper, and is an interesting direction for future research.

\subsection{Enabling Technologies}

\begin{displayquote}
\emph{The majority of required technologies are mature and ready for incorporation in a distributed instrument architecture. However, additional focused research and development are needed.}
\end{displayquote}

We surveyed the state-of-the-art in enabling technologies for distributed instruments, namely, EDL and sensor placement, localization and synchronization, communications, on-board computation and storage, instrument autonomy, power, and thermal control. For each of these technologies, we compared the state-of-the-art with the likely requirements of a distributed instrument architecture for each of the science topics identified above (Figure \ref{tab:enabling-technologies}).

Remarkably, the majority of required technologies are quite mature and ready for incorporation in a distributed instrument architecture. However, further research and development is required in three key areas:
\begin{itemize}
\item  \textbf{EDL and sensor placement techniques} that can ensure global access to planetary surfaces, tailored to specific Solar System bodies of interest;
\item \textbf{Compact and long-lasting power solutions} for sensing units, especially for high latitudes and for exploration of bodies in the outer Solar System; 
\item \textbf{Science autonomy}, especially data compression for distributed seismology and active sampling techniques to curtail power requirements for all classes of distributed instruments.
\end{itemize}

\subsection{Next Steps}

Distributed instruments are uniquely well-positioned to address a number of high priority scientific questions, and key technologies are either mature or well on their way to feasibility. To fully realize the promise of distributed instruments, we envision that additional research and development work will need to focus in two areas.

First, continued research and development in EDL and sensor placement, compact and long-lived power solutions, and instrument autonomy will be critical to fill technology gaps in these areas and bring them to a technology readiness level sufficient for inclusion in future mission proposals.

Second, while this paper provided a high-level exploration of the technology challenges inherent in distributed instruments, detailed architecture studies need to be conducted to identify scientific requirements and resolve system-level trades for future distributed instruments. For each of the  science questions identified in Section \ref{ch:sciencecase}, systems-level studies will have to quantitatively define the number of sensors carried by a specific distributed instrument, their desired placement on the surface of the target body, the required measurement frequency and overall duration of the mission, and the sensitivity of individual sensors, in order to inform the selection of relevant flight opportunities and drive follow-on technology development.

We envision that, together, these areas will bring the promise of distributed instruments closer to reality, and bring the planetary science community closer to unlocking key questions in planetary science.

\acknowledgments
This research was carried out at the Jet Propulsion Laboratory, California Institute of Technology, under a contract with the National Aeronautics and Space Administration (80NM0018D0004).

\ifanonymous
\else

The work was sponsored by a grant furnished by JPL's Office of Strategic Planning. The authors would like to thank Dr. Leon Alkalai and Dr. Adrian Stoica for their support of this work.

The authors would also like to thank
Adrian Agogino,
Greg Allen,
Nathan Barba,
Dave Berger,
Bruce Bills,
John Bodylski,
Albion Bowers,
Samuel Case Bradford,
Kar-Ming Cheung,
Steve Chien,
Keith Chin,
Lance Christensen,
Christine Gebara,
Louis Giersch,
Kris Gorski,
Laura Kerber,
Stefano Morellina,
Hiro Ono,
Mark Panning,
Ryan Pavlick,
Julian Rimoli,
Bill Smythe,
Steve Vance,
Massimo Vespignani,
William Walsh, and
Wayne Zimmerman
for the invaluable feedback and for lending their scientific and technical expertise to this study.

We are grateful to David Hinkle from the JPL Studio for the wonderful concept art.

Finally, we would like to thank the A-Team core team: 
Melle Amade, 
Mark Chodas, 
Benji Donitz, 
Karla Hawkinson, 
Valerie Scott, 
Rashmi Shah, 
Austin Tran, and
Randii Wessen,
for organizing and running an exceptional A-Team study.

\copyright 2024  California Institute of Technology. Government sponsorship acknowledged.

Pre-decisional information – for planning and discussion purposes only.
\fi

\bibliographystyle{aasjournal} %
\bibliography{
intro.bib,
history.bib,
science_case.bib,
seismology.bib,
magnetometry.bib,
trace_gas.bib,
mars_atmosphere.bib,
pnt.bib,
communications.bib,
thermal.bib,
blue_skies.bib,
autonomy.bib
}

\end{document}